\documentclass{aa}  

\usepackage{tabularx}
\usepackage{booktabs}
\usepackage{makecell}

\usepackage{tikz}
\usepackage{adjustbox}
\usetikzlibrary{arrows.meta, positioning, fit}

\tikzset{
  box/.style={rectangle, rounded corners, draw=black, fill=blue!7,
              text centered, minimum height=1.2cm, minimum width=3.2cm, font=\small},
  arrow/.style={-{Latex[scale=1.2]}, thick},
  groupbox/.style={draw=red!60, rounded corners, dashed, inner sep=0.3cm}
}

\usepackage{hyperref}
\usepackage{graphicx}
\usepackage{xcolor}
\usepackage{txfonts}
\usepackage{amsmath}
\usepackage{lipsum}
\usepackage[utf8]{inputenc}               % if needed
\usepackage[T1]{fontenc}
\usepackage{natbib}
\usepackage{ulem}

\usepackage{subcaption}         % necessary for continued figures, example in section 3
\usepackage{lscape}             % to rotate a single page table, example in appendix.
\usepackage{placeins}           % useful with \FloatBarrier, to keep 
\begin{document}

%%%%%%%%%%%%%%%%%%%%%%%%%%%%%%%%%%%%%%%%
% if you use custom commands in your title,
% ensure to check your title when submitting!
%%%%%%%%%%%%%%%%%%%%%%%%%%%%%%%%%%%%%%%%
\title{On the robustness of the angular homogeneity scale $\theta_H$: a comparative analysis of computational approaches}

   %\subtitle{}

%%%%%%%%%%%%%%%%%%%%%%%%%%%%%%%%%%%%%%%%
% Please separate each author with the \and command
%
% Please do not include ORCIDs next to author names.
% Only ORCIDs authenticated by individual authors in EDPS
% editorial system will be taken into account.
% ORCIDs included here will be removed.
%%%%%%%%%%%%%%%%%%%%%%%%%%%%%%%%%%%%%%%%

\author{P. Fanha\thanks{\email{up201908149@edu.fc.up.pt}}$^{1,2}$, A.~Da Silva$^{3,4}$, J. Fonseca$^{1,2,5}$,  J.P.~Mimoso$^{3,4}$, Z. Sakr $^{6,7,8}$}

\institute{Instituto de Astrofisica e Ci\^{e}ncias do Espa\c{c}o, Universidade do Porto CAUP, Rua das Estrelas, PT4150-762 Porto, Portugal 
\and
Departamento de Física e Astronomia, Faculdade de Ciências, Universidade do Porto, Rua do Campo Alegre, 687, PT4169-007 Porto, Portugal 
\and
Instituto de Astrof\'isica e Ci\^encias do Espa\c{c}o, Faculdade de Ci\^encias, Universidade de Lisboa, Campo Grande, PT-1749-016 Lisboa, Portugal.
\and 
Departamento de F\'isica, Faculdade de Ci\^encias, Universidade de Lisboa, Edif\'icio C8, Campo Grande, PT1749-016 Lisboa, Portugal.
\and 
Department of Physics \& Astronomy, University of the Western Cape, Cape Town 7535, South Africa
\and
Instituto de Física Teórica UAM-CSIC, Campus de Cantoblanco, 28049 Madrid, Spain
\and
Universit\'e St Joseph; Faculty of Sciences, Beirut, Lebanon 
\and
Institut de Recherche en Astrophysique et Plan\'etologie (IRAP), Universit\'e de Toulouse, CNRS, UPS, CNES, 14 Av. Edouard Belin, 31400 Toulouse, France
}

\date{\today}

% \abstract{}{}{}{}{}
% 5 {} token are mandatory
 
%%  \abstract
%%  % context heading (optional)
%%  % {} leave it empty if necessary  
%%   {Optional, leave empty if necessary.  The heading “Context” is used when needed to give background information on the research conducted in the paper}
%%  % aims heading (mandatory)
%%   {Mandatory. The objectives of the paper are defined here.} 
%%  % methods heading (mandatory)
%%   {Mandatory. The methods of the investigation are outlined here}
%%  % results heading (mandatory)
%%   {Mandatory. The results are summarized here.}
%%  % conclusions heading (optional), leave it empty if necessary
%%   {Optional, leave empty if necessary.  “Conclusions” can be used to explicit the general conclusions that can be drawn from the paper.}

\abstract
{The assumption of large-scale homogeneity is a cornerstone of modern cosmology and underlies the validity of the FLRW framework. Testing the scale at which the Universe transitions to homogeneity remains a key observational challenge, particularly with the increasing precision of galaxy surveys.}
{We aim to assess the robustness of the angular homogeneity scale, $\theta_H$, by systematically comparing different computational approaches used in its estimation and by quantifying the impact of methodological choices and physical clustering scales.}
{We analyse mock galaxy catalogues from the MICE Grand Challenge simulation. The angular fractal dimension $D_2(\theta)$ is computed using the Landy--Szalay estimator and direct pair-counting methods. We implement different approaches, including symbolic regression, to model $D_2(\theta)$ and determine $\theta_H$. Uncertainties are estimated using resampling techniques and alternative parametric error propagation methods.}
{We find that the estimation of $\theta_H$ is sensitive to methodological choices in the analysis, such as survey area, redshift bin, numerical implementation and fitting strategy. While its redshift evolution is robust, its absolute value is sensitive to both modelling choices and the presence of local clustering features.}
{Our results highlight the importance of methodological systematics in homogeneity studies, showing that the determination of $\theta_H$ depends not only on the data, but also on the adopted analysis strategy. Flexible approaches such as symbolic regression provide a useful framework to model these effects, but also emphasize the need for careful modelling and survey design. This has important implications for future large-scale structure analyses aiming to test the Cosmological Principle with high precision.}


\titlerunning{On the robustness of the angular homogeneity scale $\theta_H$: a comparative analysis of computational approaches}
\authorrunning{P. Fanha, et al.}

\maketitle

\nolinenumbers

%%%%%%%%%%%%%%%%%%%%%%%%%%%%%%%%%%%%%%%%%%%%%%%%%%%%%%%%%%%%%%
\section{Introduction}

One of the foundational assumptions of modern cosmology is the \textit{Cosmological Principle}, the idea that the Universe is statistically homogeneous and isotropic on sufficiently large scales. This principle underpins the standard $\Lambda \text{CDM}$ model based on the Friedmann-Lemaître-Robertson-Walker (FLRW) metric and plays a pivotal role in the interpretation of a wide range of cosmological observations. While the statistical isotropy of the Universe has been generally confirmed, most notably through the analysis of the Cosmic Microwave Background (CMB), direct tests of spatial homogeneity remain more subtle and challenging; see \cite{Aluri_2023} and references therein.

To probe homogeneity in the large-scale distribution of matter, statistical tools derived from galaxy clustering are commonly used. Among these, the fractal dimension $D_2$ has emerged as a powerful indicator of the scale at which the Universe transitions from a clustered to a homogeneous distribution. When combined with the concept of the three-dimensional homogeneity scale $R_H$ or its angular counterpart $\theta_H$, these measures offer a direct observational way to test the scale of homogeneity across redshifts and assess the validity of the Cosmological Principle.

Historically, the study of cosmic homogeneity began with three-dimensional analyses to determine the spatial fractal dimension $D_2(r)$. Foundational work by \cite{Hogg_2005} demonstrated homogeneity using Luminous Red Galaxies, followed by \cite{Scrimgeour_2012}, which used the WiggleZ Dark Energy Survey to identify a transition to homogeneity scale, $R_H$, of approximately $70~h^{-1}\,\text{Mpc}$. Subsequent studies, such as those by \cite{Ntelis_2017} and \cite{dias2023probingcosmichomogeneitylocal}, confirmed $R_H$ to exist within the range of $60 - 80 ~ h^{-1} \, \text{Mpc}$. However, as noted in this work, these 3D measurements are inherently model-dependent, as they require the assumption of a cosmological model to convert observable redshifts into spatial distances. With this mind, our work focuses on the projected (angular) distribution of galaxies in two dimensions and the angular fractal dimension.

The literature has followed two main paths for analyses using the angular fractal dimension: a \textit{direct} method, measuring clustering and $D_2$ directly from galaxy counts-in-cells, and an \textit{indirect} method, which derives the fractal dimension from the 2PACF. Early applications to surveys such as SDSS and ALFALFA (see \cite{Goncalves_2018,Avila_2018,Avila_2019}) established the standard procedure for identifying the homogeneity scale, $\theta_H$, laying the groundwork for using angular clustering statistics not only to test the transition to homogeneity, but also to explore how this transition may evolve with redshift.

More recent work has improved the statistical reliability and cosmological utility of these measurements. \cite{andrade2022angularscalehomogeneitysdssiv} introduced a more rigorous approach to uncertainty by using bootstrap resampling of the survey footprint itself, allowing for internal error estimation without relying on external simulations. Most recently, \cite{Shao_2025} demonstrated that these measurements are not only consistency checks of the FLRW paradigm; by combining $\theta_H$ with other cosmological data, they showed that the transition to homogeneity can be used to constrain cosmological parameters, including dark energy properties, thereby bridging the gap between statistical geometry and fundamental physics.

In parallel with these developments, machine learning techniques have increasingly been explored in cosmology as flexible tools for modelling complex, non-linear phenomena. Among these, symbolic regression (SR) has emerged as a promising approach for reconstructing cosmological functions and providing accurate fitting formulae that can significantly reduce the computational cost of numerically intensive calculations (see e.g.~\citep{Arjona2020b,Aizpuru_2021,Nesseris2022_euclid, Carvalho2023_sigma8,Carvalho_2026_closedforms}). Unlike traditional parametric fitting, SR does not assume a predefined functional form, but instead searches the space of mathematical expressions to identify optimal representations of the data.

In this work, we extend the use of symbolic regression to the analysis of fractal dimension profiles $D_2(\theta)$, where, to the best of our knowledge, it has not been previously applied. This approach allows for increased functional flexibility in modelling the $D_2(\theta)$ curves, enabling the capture of local features and small-scale variations associated with galaxy clustering and observational effects. At the same time, it provides analytic expressions that facilitate the determination of the homogeneity scale and its associated uncertainties. This makes SR a useful framework for assessing the robustness of $\theta_H$ estimators, particularly in the context of next-generation wide-field galaxy surveys.

Our goal is not only to measure $\theta_H$, but also to quantify the methodological uncertainties associated with its estimation. This is particularly relevant in the context of current and upcoming wide-field galaxy surveys, where percent-level systematics can significantly affect cosmological interpretations. By comparing alternative estimators and computational strategies, we aim to identify robust and reliable procedures for testing the validity of the FLRW paradigm.

This paper is structured as follows. In Section~\ref{sc:method}, we review the theoretical background of cosmological fractal dimension analysis, the application of this analysis to data and the computational methods proposed to do so. Section~\ref{sc:data} describes the simulated data sets we consider here. In Section~\ref{sc:results}, we present the results of our different computational approaches. We conclude with a discussion of the implications of our findings in Section~\ref{sc:conclusions}.

%%%%%%%%%%%%%%%%%%%%%%%%%%%%%%%%%%%%%%%%%%%%%%%%%%%%%%%%%%%%%%

\section{Theoretical framework} \label{sc:theory}

Within the Standard Model of Cosmology, the assumption of a spatially homogeneous and isotropic Universe leads to the Friedmann-Lemaître-Robertson-Walker (FLRW) metric. Direct probes of the large-scale structure provide an independent data-driven approach to testing this homogeneity.

We adopt a cosmological model-independent approach by examining the projected distribution of objects in the sky, specifically in \textit{two} dimensions. This method utilizes a counts-in-spherical-caps analysis: for a spherical cap of angular radius $\theta$, the number of objects in the observational catalogue is denoted as $N(< \theta)$. For a theoretically homogeneous distribution, the expected count is:
\begin{equation}
    N_R(< \theta) = 2\pi\bar{\rho}(1 - \cos{\theta})
\end{equation}
where $\bar{\rho}$ represents the mean number density of pairs of objects in the sky. We define the scaled counts-in-spherical-caps as:
\begin{equation}
    \mathcal{N}(< \theta) \equiv \frac{N(< \theta)}{N_R(< \theta)}
\end{equation}
The fractal dimension in two dimensions, $D_2(\theta)$ is expressed as:
\begin{equation}
    D_2(\theta) \equiv \frac{d \, \ln{N(< \theta)}}{d \, \ln{\theta}} = \frac{\theta \sin{\theta}}{1 - \cos{\theta}} + \frac{d \, \ln{\mathcal{N}(< \theta)}}{d \, \ln{\theta}}
    \label{eq:d2-from-scaled-number-counts}
\end{equation}
For a distribution that is homogeneous at sufficiently large scales, $D_2(\theta)$ asymptotically approaches $D_2^H$:
\begin{equation}
    D_2^H \equiv \frac{\theta \sin{\theta}}{1 - \cos{\theta}}
    \label{D2_hom}
\end{equation}
This profile serves as a statistical test of the transverse (angular) homogeneity of the Universe.

\subsection{Connection to the 2PACF}
Following \cite{10.23943/princeton/9780691209814.001.0001}, the conditional probability (or differential number counts) of finding a random object within a solid angle element $\delta \Omega$ at an angular separation $\theta$ from a specific object is given by:
\begin{equation}
    \delta N = \bar{\rho} \, \delta\Omega \, [1 + \omega(\theta)]\, ,
\end{equation}
where $\omega$ is the two-point angular correlation function (2PACF) and $\bar{\rho}$ is the average density of pairs of objects. By integrating $\delta N$ over a spherical cap, we obtain the expected number of pairs within radius $\theta$:
\begin{equation}
    N(< \theta) = \rho \int \, d\Omega \, [1 + \omega(\theta)]
\end{equation}
This formulation implicitly accounts for counts over the entire survey volume. This leads to the expression for scaled counts:
\begin{equation}
    \mathcal{N}(< \theta) = 1 + \frac{1}{1 - \cos{\theta}} \int_0^{\theta} \omega(\theta') \sin{\theta'} \, d\theta'
    \label{eq:scaled-number-counts}
\end{equation}
Consequently, the fractal dimension becomes:
\begin{equation}
    D_2(\theta) = \frac{\theta \sin{\theta}}{1 - \cos{\theta}} + \frac{d}{d \, \ln{\theta}} \ln\left[1 + \frac{1}{1 - \cos{\theta}} \int_0^{\theta} \omega(\theta') \sin{\theta'} \, d\theta' \right]
    \label{eq:d2}
\end{equation}
Because the correlation function quantifies matter clustering, its form is heavily dependent on the cosmological model and redshift.

The fractal dimension profile $D_2(\theta)$ allows a clear definition of the angular homogeneity scale: the scale at which $D_2(\theta)$ equals $D_2^H(\theta)$, the homogeneity threshold curve. In practice, this may never happen due to observational artifacts. A commonly used criterion in the literature is the so-called 1\% criterion \citep{Scrimgeour_2012,Gon_alves_2017,Avila_2018,Avila_2019, andrade2022angularscalehomogeneitysdssiv, dias2023probingcosmichomogeneitylocal, shao2025cosmologicalconstraintsangularhomogeneity}, which defines the homogeneity scale $\theta_H$ as the angular radius at which $D_2(\theta)$ \textit{first} reaches 1\% from the homogeneous threshold value:
\begin{equation}
    D_2(\theta_H) = 0.99 D_2^H(\theta_H)
\end{equation}
While this criterion is somewhat arbitrary, it has the advantage of being survey-independent and therefore facilitates comparison across different datasets and against theoretical expected values. Alternative definitions have been proposed in the literature \citep{Hogg_2005, Heinesen_2020}, though in this work we adopt the 1\% criterion and leave it as an open question whether it is the most adequate to use in future work.

\subsection{Numerical estimators} \label{ssec:num_est}
The computation of $\omega(\theta)$, $\mathcal{N}(< \theta)$, and $D_2(\theta)$ from observational data, involves the use of 
statistical estimators based on pair counts. For a catalogue containing binned redshift data with the right ascension and declination of each object, we determine $DD(\theta)$ by counting object pairs separated by an angular distance $\theta$. We emulate a homogeneous distribution (a random catalogue) by placing objects randomly on the spherical surface considering the original survey footprint, to yield the pair counts $RR(\theta)$. We also account for cross-pairs between the data and the random catalogue, denoted as $DR(\theta)$.

Several estimators for the two-point angular correlation function (2PACF) have been developed from these quantities. Among these, the Landy-Szalay estimator (Eq.~(\ref{eq:ls-estimator-wtheta})) is widely regarded as the most robust \citep{Hirata2009_CF} and is the choice adopted in this work.
\begin{equation}
    \omega_{LS}(\theta) = \frac{DD(\theta) - 2DR(\theta) + RR(\theta)}{RR(\theta)}
    \label{eq:ls-estimator-wtheta}
\end{equation}
In all instances, the pair counts $DD$, $DR$, and $RR$ are normalized by the total number of pairs in their respective catalogues. This normalization ensures the resulting correlation function is independent of the absolute number of objects, avoiding inherent biases in catalogues of differing sizes.

The scaled counts can be expressed in terms of the Landy-Szalay estimator as:
\begin{equation}
    \mathcal{N}_{LS}(< \theta) = 1 + \frac{1}{1 - \cos{\theta}} \int_0^{\theta} \omega_{LS}(\theta') \, \sin{\theta'} \, d\theta'
\end{equation}
Alternatively, we can employ a direct approach by summing pair counts explicitly \citep[see e.g.][]{fanha2025angularscale}:
\begin{equation}
    \mathcal{N}_{LS}(< \theta) = 1 + \frac{\sum_{\theta_i = 0}^{\theta} [DD(\theta_i) - 2DR(\theta_i) + RR(\theta_i)]}{\sum_{\theta_i = 0}^{\theta} RR(\theta_i)}
    \label{eq:ls-estimator-scaled-number-counts}
\end{equation}

%%%%%%%%%%%%%%%%
\section{Methodology} \label{sc:method}

\subsection{Computational methodology and uncertainty estimation}
While previous works have provided valuable first measurements, they relied on samples with modest galaxy counts and incomplete sky coverage. Next-generation surveys will dramatically improve this landscape. These datasets necessitate a more rigorous statistical framework to determine the angular homogeneity scale, $\theta_H$, and its associated uncertainties.

In general, the analysis follows one of two paths:
\begin{itemize}
    \item \textbf{Indirect Method:} Measuring the 2PACF via pair counts (Eq.~\ref{eq:ls-estimator-wtheta}) and using it to derive the fractal dimension profile $D_2(\theta)$ (Eq.~\ref{eq:scaled-number-counts} and Eq.~\ref{eq:d2}).
    \item \textbf{Direct Method:} Measuring the fractal dimension curve directly from pair counts (Eq.~\ref{eq:ls-estimator-scaled-number-counts} and Eq.~\ref{eq:d2-from-scaled-number-counts}).
\end{itemize}
The scale $\theta_H$ is defined as the intersection of the $D_2(\theta)$ curve obtained with a 1\% homogeneity threshold.

The indirect approach suffers from one limitation: although obtaining the angular correlation function is more straightforward and one may smooth out the unwanted small-scale oscillations (see e.g.~\citep{Avila_2018}), it assumes a perfect, regular footprint for the random distribution, which might not be the case when applying the method to observational data. \\

We utilize the Landy-Szalay (LS) estimator, with pair counts calculated using the Python package {\tt TreeCorr} at logarithmic-spaced angular separations $\theta$. The number of angular separation bins is determined heuristically to accurately model the $D_2(\theta)$ curve while minimizing oscillations due to the discrete nature of the estimator. For upcoming surveys, the variance introduced by random catalogues is negligible compared to sample variance. To estimate observational uncertainty, we adopt a $\omega(\theta)$ or $D_2(\theta)$ covariance matrix-based approach derived directly from survey data by subdividing the area into spatial patches, taking advantage of the built-in features in {\tt TreeCorr}. We follow the weighting schemes proposed by \citep{Mohammad_2022} for jackknife and bootstrap resampling. These schemes improve covariance estimation by down-weighting, rather than discarding, information from galaxy pairs crossing patch boundaries. To ensure statistical independence, the typical patch size must exceed the maximum correlation length inferred from $\omega(\theta)$. We determine the ideal number of patches by checking for the convergence of the covariance matrix structure.

Previous studies utilized fixed functional-form fitting, which often fails to reproduce the proper shape of $D_2(\theta)$ across a wide interval of redshift bins. We propose an alternative approach, where we replace these with \textit{Symbolic Regression} (SR), an unsupervised machine learning method that searches the space of mathematical expressions to identify the best fit without predefined parameters. Specifically, given a user-defined list of operators and functions, the building blocks of the mathematical expressions, SR evolves populations of candidate expressions using natural-selection-inspired rules, optimizing both fit accuracy and expression simplicity.

\begin{figure}
    \centering
    \adjustbox{max width=0.5\textwidth}{ 
        \begin{tikzpicture}[node distance=0.5cm]
            % Main pipeline nodes
            \node[box] (wtheta) {\shortstack{Fit to $\omega(\theta)$ LS \\ (mean from all randoms) \\ using SR}};
            \node[box, right=of wtheta] (srparam) {\shortstack{Parametrize SR fit \\ $\omega_{\text{fit}} = \omega_{\text{fit}}(\theta;A, B, C, \dots)$}};
            \node[box, right=of srparam] (ncounts) {\shortstack{Analytical $\mathcal{N}(< \theta)$ \\ plugging in previous SR fit}};
            \node[box, below=of ncounts] (d2def) {\shortstack{$D_2$ \\ using definition \\ (analytical calculation \\ using {\tt SymPy\footnote{https://www.sympy.org}})}};
            \node[box, left=of d2def] (paramunc) {\shortstack{Determine \\ fit parameters' uncertainties}};
            \node[box, left=of paramunc, fill=green!15] (final) {\shortstack{Final measurement: \\ dependent on \\ uncertainty estimation \\ method}};
            
            % Arrows
            \draw[arrow] (wtheta) -- (srparam);
            \draw[arrow] (srparam) -- (ncounts);
            \draw[arrow] (ncounts) -- (d2def);
            \draw[arrow] (d2def) -- (paramunc);
            \draw[arrow] (paramunc) -- (final);
        \end{tikzpicture}
    }
    \caption{Flow diagram of the SR approach using the \textbf{indirect} method to estimate the angular homogeneity scale and uncertainty.}
    \label{fig:flow_thetaH_estimation_SR_indirect}
\end{figure}
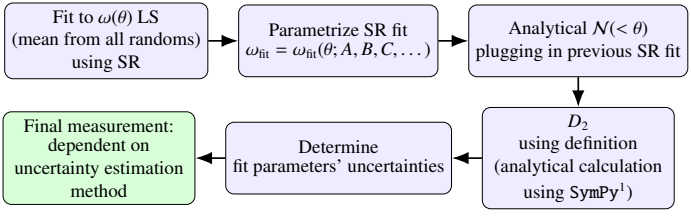

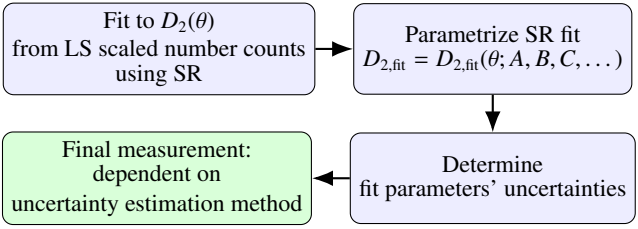
\begin{figure}
    \centering
    \adjustbox{max width=0.5\textwidth}{ 
        \begin{tikzpicture}[node distance=0.5cm]
            % Main pipeline nodes
            \node[box] (d2) {\shortstack{Fit to $D_2(\theta)$ \\ from LS scaled number counts \\ using SR}};
            \node[box, right=of d2] (srparam) {\shortstack{Parametrize SR fit \\ $D_{2,{\text{fit}}} = D_{2,\text{fit}}(\theta;A, B, C, \dots)$}};
            \node[box, below=of srparam] (paramunc) {\shortstack{Determine \\ fit parameters' uncertainties}};
            \node[box, left=of paramunc, fill=green!15] (final) {\shortstack{Final measurement: \\ dependent on \\ uncertainty estimation method}};
            
            % Arrows
            \draw[arrow] (d2) -- (srparam);
            \draw[arrow] (srparam) -- (paramunc);
            \draw[arrow] (paramunc) -- (final);
        \end{tikzpicture}
    }
    \caption{Flow diagram of the SR approach using the \textbf{direct} method to estimate the angular homogeneity scale and uncertainty.}
    \label{fig:flow_thetaH_estimation_SR_direct}
\end{figure}

The advantages of SR include:
\begin{itemize}
    \item \textbf{Interpretability}: Produces unbiased/data-driven closed-form expressions with no free parameters that can be compared to theoretical expectations;
    \item \textbf{Flexibility}: Does not require preset functional forms;
    \item \textbf{Uniform Quality}: Guarantees fits across bins via predefined quality thresholds.
\end{itemize}
For our direct method, we apply SR to the $D_2(\theta)$ LS data points using basic operators ($+$, $-$, $\times$, $\div$, $\exp$) and set complexity at an arbitrary, large value (e.g. $40$). The fitness function is defined as the \textbf{weighted mean squared error} (WMSE), using the inverse covariance matrix as weights. The stopping criterion is set at $\text{WMSE} \le 1$, meaning residuals are at or below the measurement noise (as such, any other fit that satisfies this condition is equally a good fit and the different expressions found cannot be distinguished). The selection of the best SR candidate expression is determined as a balance of complexity and accuracy (as given by the {\tt PySR} parameter {\tt model\_selection = 'best'}). The two paths, indirect and direct method, can thus be summarized via the two diagrams, Fig.~\ref{fig:flow_thetaH_estimation_SR_indirect} and Fig.~\ref{fig:flow_thetaH_estimation_SR_direct}, respectively. \\

\textbf{Uncertainty in $\theta_H$}:
To estimate the uncertainty of $\theta_H$, we reparametrize the SR expression, replacing constants with variables/parameters while maintaining the mathematical structure, as described in~\citep{fanha2025angularscale}, and employ a linear approximation method to estimate the covariance of those parameters relative to the covariance in the data as estimated by the jackknife/resampling resampling techniques mentioned before. The covariance of those parameters is then propagated to $\theta_H$. \\

\textbf{Covariance stability and regularization}.
The cumulative nature of $N(< \theta)$ makes $D_2(\theta)$ highly sensitive to correlation structures, often resulting in covariance matrices with large condition numbers $\kappa$. A high $\kappa$ results in numerical instability, where matrix inversion, used to calculate the WMSE, amplifies errors. To mitigate this, we apply regularization of the covariance matrix $C$ by diagonal loading:
\begin{equation}
    C' = C + \lambda I
\end{equation}
The regularization parameter $\lambda$ is chosen such that the new covariance matrix $C'$ is structurally similar to $C$ and its smallest eigenvalues differ only by a very small value, enough to stabilize matrix inversion but never result in underestimating uncertainties in $\theta_H$. The value of $\lambda$ is selected as the maximum value provided by two constraints:
\begin{enumerate}
    \item \textbf{Condition Number Constraint}: Ensuring $\kappa(C) \le \kappa_{target}$, where:
        \begin{equation}
            \lambda = \frac{\lambda_{max} - \kappa_{target} \lambda_{min}}{\kappa_{target} - 1}
        \end{equation}
    \item \textbf{Variance Preservation}: Ensuring $\lambda$ does not exceed a fraction $f$ of the mean diagonal variance: $\lambda = f \langle \text{diag} (C) \rangle$
\end{enumerate}
Throughout this work, we adopt $\kappa_{target} = 10^4$ and $f = 0.01$ for all calculations.

%%%%%%%%%%%%%%%%%%%%%%%%%%%%%%%%%%%%%%%%%%%%%%%%%%%%%%%%%%%%%%
\section{Data} \label{sc:data}

\subsection{Galaxy mock Simulations \label{sec:sim_data}}
We use ready-made galaxy mock simulations derived from the Mice Grand Challenge (MICE-GC) N-body simulation, designed to model the large-scale structure of the Universe assuming large-scale homogeneity and isotropy (FLRW cosmology). These simulations provide estimates of galaxy luminosities tailored for wide-field surveys such as the Dark Energy Survey (DES) \cite{Abbot_2021} and the Euclid mission \cite{Scaramella-EP1}. The mock galaxy catalogue is constructed using a hybrid approach that combines a Halo Occupation Distribution (HOD) (see e.g. \cite{Jing_2002,2001ApJ...546...20S,Berlind_2002}) model with Halo Abundance Matching (HAM) (see e.g. \cite{2004MNRAS.353..189V,Tasitsiomi_2004,2006ApJ...647..201C}) to populate Friends-of-Friends (FoF; see e.g. \cite{1982ApJ...257..423H,1982ApJ...259..449P,1984MNRAS.206..559T}) dark matter halos identified in the MICE-GC simulation. 
The simulation assumes a flat $\Lambda$CDM cosmology with matter density parameter $\Omega_m = 0.25$, dark energy density $\Omega_\Lambda = 0.75$, baryon density $\Omega_b = 0.044$, amplitude of matter fluctuations on scales of $8 \, h^{-1} \, \text{Mpc}$  $\sigma_8 = 0.8$, scalar spectral index $n_s = 0.95$, and dimensionless Hubble parameter $h = 0.7$.

The full version 2 of the MICE mock galaxy catalogue \cite{Fosalba_2015a,Crocce_2015,Fosalba_2015b} comprises 499,609,997 galaxies, making it one of the largest currently in existence, and is publicly available through the CosmoHub\footnote{\href{https://cosmohub.pic.es}{CosmoHub platform: https://cosmohub.pic.es}} portal \citep{Carretero_2017}. For our analysis, we use a smaller subset of the catalogue, covering roughly 3,438 square degrees from a sky region where DES i-band magnitudes are complete down to $i = 24$, and Euclid H-band magnitudes are complete down to $H \sim 23.0$, up to $z \sim 1.4$, though we select only galaxies up to $i = 23.5$. 

We bin the galaxy catalogue in redshift and select bins with left edges at $z = 0.3$, $0.5$, $0.7$, $0.9$, and $1.1$, each with a width of $\Delta z = 0.01$. 
These redshift bins, together with the adopted magnitude cut and bin width $\Delta z = 0.01$, were chosen so as to ensure good counting statistics in all slices, as indicated in Table~\ref{tab:redshift_counts}. The choice of a narrow and fixed redshift width is also motivated by the need to control projection effects in the angular correlation function, which are known to affect the determination of the BAO angular scale (see e.g.~\citealt{Sanchez_2011}). In this context, the BAO angular scale is defined as
\begin{equation}
    \theta_{\mathrm{BAO}}(z) = \frac{r_s}{(1+z)\,D_A(z)},
\end{equation}
where $r_s$ is the sound horizon at decoupling and $D_A(z)$ is the angular diameter distance at redshift $z$ (see e.g.~\citealt{Aizpuru_2021}). For the present choice of redshift slices and for the MICE cosmology adopted here, $\theta_{\mathrm{BAO}}$ takes values that are close to, and in some cases overlap with, the expected homogeneity scale $\theta_H(z)$ over the redshift range considered. This choice is intentional: it allows us to assess how robust the different estimation methods remain when the angular BAO scale lies near the transition-to-homogeneity scale. More broadly, testing the sensitivity of the methodology under such potentially challenging conditions is one of the goals of this work.

Table~\ref{tab:redshift_counts} presents the average redshift and number of objects in each redshift bin we use in this paper.
\begin{table}
\centering
\begin{tabular}{|c|c|c|}
\toprule
\textbf{Redshift bin} & \textbf{Average redshift} & \textbf{Number of objects} \\
\midrule
0.3--0.31 & 0.305 & 1,543,290 \\
0.5--0.51 & 0.505 & 1,895,610 \\
0.7--0.71 & 0.705 & 1,938,871 \\
0.9--0.91 & 0.905 & 1,652,408 \\
1.1--1.11 & 1.105 & 1,105,749 \\
\bottomrule
\end{tabular}
\caption{Average redshift and number of objects in each redshift bin for the MICECAT catalogue within the entire completeness region used in this work.}
\label{tab:redshift_counts}
\end{table}

%thesis-regions-600-1200-z=0.3.png
\begin{figure}
    \centering
    \includegraphics[width=\linewidth]{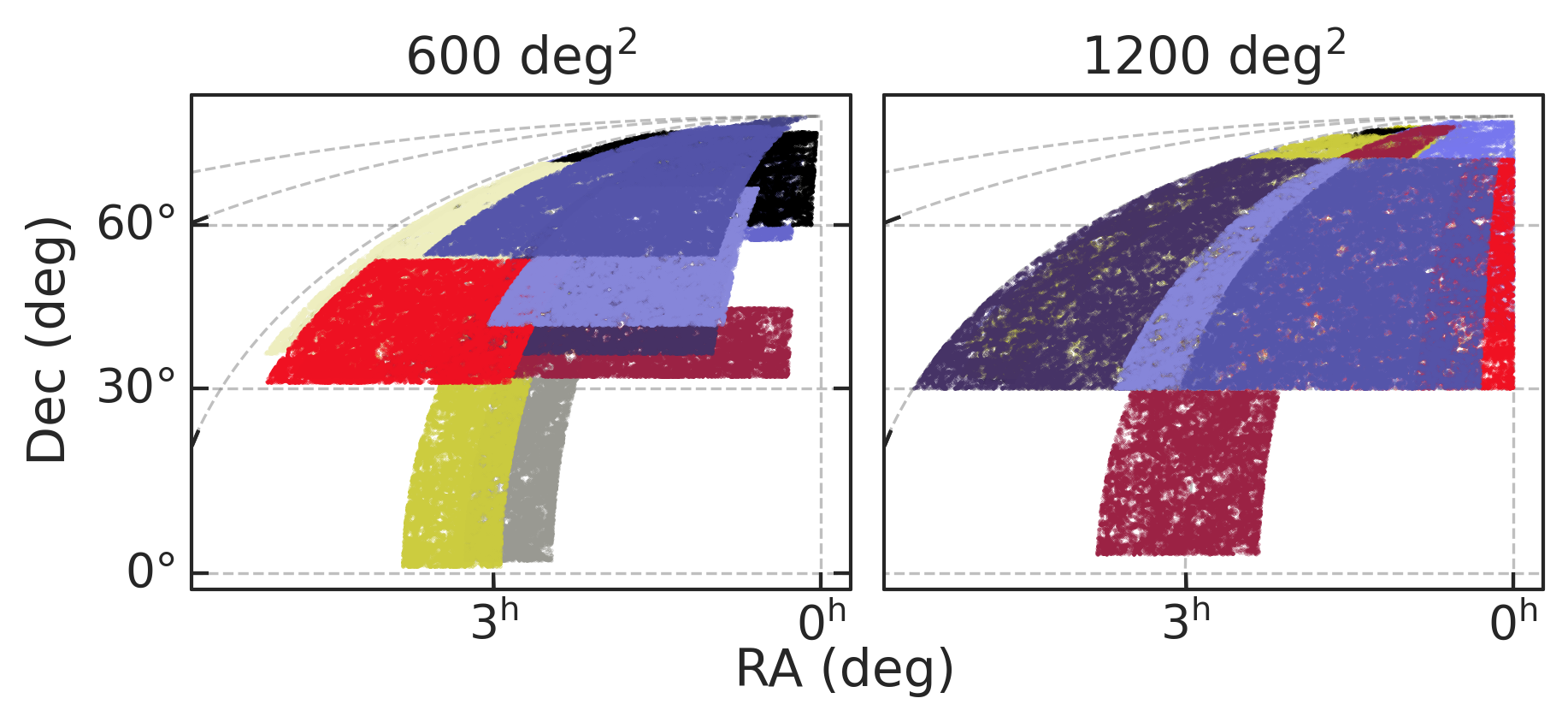}%,height=5cm,trim=0.4cm 0.2cm 0cm 0cm,clip]
    \caption{Distribution on the sky of the 16 regions of $600~\text{deg}^2$ (left panel) and of the 16 regions of $1200~\text{deg}^2$ (right panel) selected from MICECAT within the completeness region (example for $z=0.3$). Because of the overlap between regions, many are partially or completely hidden.}
    \label{fig:survey-regions-map-600-vs-1200-z=0.3}
\end{figure}

\subsection{Data analysis procedure}
For all analyses, we utilise 30 log-spaced angular separation bins ranging from $0.5^\circ$ to $20^\circ$. To emulate a homogeneous baseline, we generate 20 random catalogues for each survey region, each containing the same number of objects as the data catalogue. \\

%thesis-patches-subsampling-1200-z=0_3.png
\begin{figure}
    \centering
    \includegraphics[width=0.875\linewidth,height=6.6cm,keepaspectratio=false]{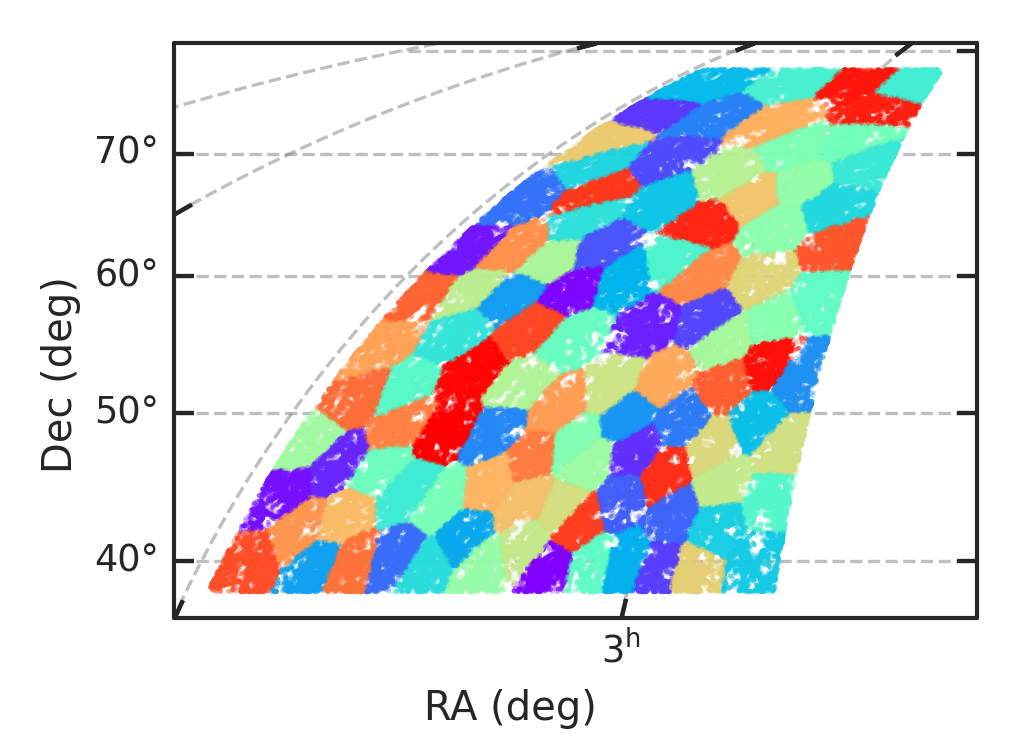}
    \caption{Example of region subsampled into 100 roughly equal-area patches by {\tt TreeCorr} for uncertainty estimation.}
    \label{fig:patches-subsampling-example}
\end{figure}

\textbf{Survey subpartitioning}.
To analyse how survey area influences $\theta_H$ and to assess variability across different sky patches, we subpartition the full catalogue as defined above into multiple, smaller rectangular regions (RA, Dec) with no irregularities or holes. We select (see Fig. \ref{fig:survey-regions-map-600-vs-1200-z=0.3}):
\begin{itemize}
    \item \textbf{16 regions} of $600\,\text{deg}^2$;
    \item \textbf{16 regions} of $1200\,\text{deg}^2$;
    \item \textbf{One region} covering the full survey area ($3400\,\text{deg}^2$).
\end{itemize}
While these subpartitions overlap due to practical limits of the survey footprint, they allow us to mimic the scale of near-future surveys. We note that while differences in region shape could theoretically affect $\theta_H$, our preliminary tests suggest these effects are negligible; however, these findings are not formally presented and should be interpreted with caution. \\

\textbf{Patch generation and resampling}.
To estimate uncertainties on $\theta_H$, we further subdivide each survey region into patches using the \textit{K-means clustering algorithm} built into {\tt TreeCorr}. This algorithm produces patches of roughly equal area.

The number of patches is subject to two competing constraints:
\begin{enumerate}
    \item \textbf{Lower limit}: To ensure enough statistical samples of $\omega(\theta)$ and $D_2(\theta)$, jackknife/bootstrap resampling needs at least as many patches as angular separation bins;
    \item \textbf{Upper limit}: Each patch must be large enough so its typical scale exceeds the correlation length of the region, ensuring the statistical independence required for resampling.
\end{enumerate}
We vary the number of patches up to \textbf{200} for the full-area region, staying safely below the upper limit while maximizing the resolution of within-region variability. In Figure \ref{fig:patches-subsampling-example} we plot an example of the patch divided into 100 equal areas. Our observations indicate:
\begin{itemize}
    \item \textbf{Full-Area region}: The covariance structure stabilizes once $N_{patch} \ge 100$. To ensure maximum robustness while remaining below the physical correlation limit, we adopt $N_{patch} = 200$ for this region;
    \item \textbf{$1200\,\text{deg}^2$ regions}: Stability is reached at approximately 100-150 patches;
    \item \textbf{$600\,\text{deg}^2$ regions}: We selected 50 patches to maintain a typical scale in a patch smaller than the correlation length of the survey.
\end{itemize}

We compared bootstrap and jackknife resampling. While both methods estimate the covariance matrix with sufficient accuracy, we adopt jackknife resampling as the preferred method for all subsequent analyses due to being:
\begin{itemize}
    \item \textbf{more computationally efficient}: jackknife is significantly faster as the number of iterations always equals the number of patches, whereas bootstrap requires a much higher number of iterations to converge;
    \item \textbf{physically consistent}: jackknife does not generate physically unrealistic resampled regions.
\end{itemize}

\textbf{Selection of the method for calculating $D_2$}.
Following our comparative testing of both analytical paths, we have opted to utilize the direct method for the remainder of this work. This choice was motivated by the fact that its implementation is simpler and less prone to numerical instability and that it does not suffer from the limitation described in Section~\ref{sc:method} regarding the theoretical prediction for random counts.

%%%%%%%%%%%%%%%%%%%%%%%%%%%%%%%%%%%%%%%

\section{Results} \label{sc:results}

\subsection{Angular correlation and $D_2(\theta)$ determination}

\begin{figure}
    \centering
    \includegraphics[width=\linewidth,trim=0.3cm 0.4cm 0cm 0cm,clip]{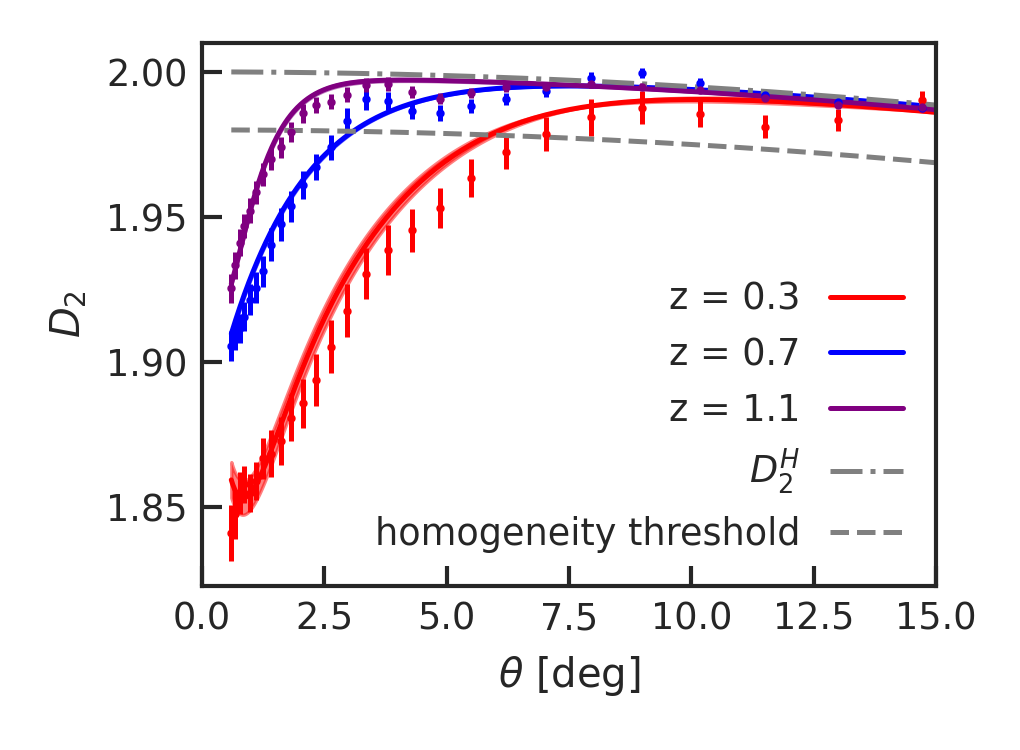}
    \caption{{Representative symbolic-regression reconstructions of the profiles $D_2(\theta)$ for three redshift bins in the full-area MICECAT region. Points with error bars correspond to the measured profiles, solid curves with a shaded region around show the symbolic-regression fits and the estimated confidence interval, and the horizontal threshold marks the homogeneity criterion.}}
    \label{fig:d2-analytic-profiles}
\end{figure}

Fig.~\ref{fig:d2-analytic-profiles} shows three representative profiles of $D_2(\theta)$ obtained from the direct estimator together with their symbolic-regression reconstructions. For clarity, only the redshift bins $z=0.3$, $z=0.7$, and $z=1.1$ are displayed, spanning the low-, intermediate-, and high-redshift regimes probed in this work. 

%D2 profiles and SR reconstructions

The corresponding analytical expressions for all redshift bins are listed in Table~\ref{tab:d2-analytic-profiles}, where we report the reconstructed forms for $D_2^*(\theta)=D_2(\theta)-0.99D_2^H(\theta)$ and their mean squared errors. The symbolic-regression models were selected using the \texttt{model\_selection='best'} criterion in PySR, which identifies the expression lying closest to the optimal trade-off between accuracy and complexity along the Pareto front. This selection avoids overfitting while retaining sufficient flexibility to capture the relevant features of the data.

\begin{table}
\centering
\resizebox{\columnwidth}{!}{
\begin{tabular}{l c c l}
\toprule
\textbf{Redshift} & \textbf{MSE} & \textbf{Expression for $D_2^*(\theta)$} & \textbf{Parameters} \\
\midrule
$z = 0.3$ & $8.07 \times 10^{-5}$ & $\displaystyle a_0 + a_1 \left( a_2 \theta + a_3 \theta \exp({\theta}) \right) \exp({a_4 \theta})$ & \makecell{$a_0 = 0.0175$ \\ $a_1 = 0.1170$ \\ $a_2 = -3.1485$ \\ $a_3 = -1.1831$ \\ $a_4 = -1.6539$}\\

\midrule
$z = 0.5$ & $4.29 \times 10^{-5}$ & $\displaystyle a_0 + a_1 a_2^{\theta} \exp({2\theta})$ & \makecell{$a_0 = 0.0195$ \\ $a_1 = -0.1722$ \\ $a_2 = 0.0776$} \\

\midrule
$z = 0.7$ & $2.13 \times 10^{-5}$ & $\displaystyle a_0 + a_1 \exp({a_2 \theta})$ & \makecell{$a_0 = 0.0192$ \\ $a_1 = -0.1297$ \\ $a_2 = -0.6172$} \\

\midrule
$z = 0.9$ & $8.69 \times 10^{-6}$ & $\displaystyle a_0 + a_1 (a_2 + \theta) \exp({-\theta})$ & \makecell{$a_0 = 0.0193$ \\ $a_1 = -0.0498$ \\ $a_2 = 2.1790$} \\

\midrule
$z = 1.1$ & $9.07 \times 10^{-6}$ & $\displaystyle a_0 + \frac{a_1 \theta \exp({a_3 \theta})}{a_2 - \theta}$ & \makecell{$a_0 = 0.0182$ \\ $a_1 = 0.5498$ \\ $a_2 = -0.9400$ \\ $a_3 = -1.8319$} \\

\bottomrule
\end{tabular}
}
\caption{{Mean squared error and symbolic regression expressions for $D_2^*(\theta) = D_2(\theta) - 0.99D_2^H(\theta)$ in different redshift bins.}}
\label{tab:d2-analytic-profiles}
\end{table}

At small angular scales, $D_2(\theta)$ departs significantly from homogeneity, reflecting the clustered nature of the galaxy distribution, while at larger angular separations it gradually approaches the homogeneous limit $D_2^H(\theta)$ given by Eq.~(\ref{D2_hom}). Superimposed on this global trend, the observed profiles exhibit local fluctuations, which may arise from genuine structure in the matter distribution, such as collapsed systems and voids, but also from survey-related effects, including incomplete sky coverage and the imprint of BAO-scale clustering.

In this context, symbolic regression offers an important advantage over standard parametric fitting functions. Rather than imposing a fixed functional form, it allows the data itself to determine a flexible analytic representation of the profile. This added flexibility enables the fit to absorb fluctuations in strongly inhomogeneous regimes ($\theta < \theta_H$) and in scales already approaching homogeneity ($\theta \gtrsim \theta_H$), while preserving a stable description of the transition region itself. As a result, the inferred intersection with the homogeneity threshold is less sensitive to local distortions of the $D_2(\theta)$ curve than in conventional global parametric fits, and avoids the limitations of local fits restricted to the neighbourhood of the intersection, which can themselves be strongly affected by fluctuations at those angular scales.

The reconstructed expressions listed in Table~\ref{tab:d2-analytic-profiles} also show that the symbolic-regression solutions remain relatively compact, while adapting their complexity to the structure of each redshift bin. In particular, the higher-redshift profiles are captured by simpler saturating forms, whereas the lower-redshift cases require more structured expressions, consistent with the stronger clustering and increased profile irregularity expected at later cosmic times. These functions should not be interpreted as fundamental physical laws, but rather as effective analytic representations of $D_2^*(\theta)$ for the MICECAT simulation and cosmology adopted here.

\subsection{Uncertainty estimation}
\label{subsc:uncertainty-estimation}

A key aspect of the determination of the homogeneity scale $\theta_H$ is the robust estimation of its associated uncertainty. In this work, we compare three different approaches (an in-depth description can be found in \cite{fanha2025angularscale}):

\begin{enumerate}
    \item A linear approximation based on error propagation, \label{enum:case-1-uncertainty}
    \item A nested sampling method applied to the symbolic regression (SR) parametrisation, \label{enum:case-2-uncertainty}
    \item And a third, parametric, resampling-based approach, inspired by existing methods in the literature, where: 
        \begin{enumerate}
            \item From jackknife resamples of the survey region, we generate an array of $D_2$ profiles, describing the intrinsic variability of $D_2$ within that region,
            \item We generate the distribution of $\theta_H$ values for this region by fitting a univariate spline over each of the $D_2$ profiles and interpolating to find the angular homogeneity scale,
            \item We estimate the most likely $\theta_H$ value and the uncertainty interval as the 16/84\% percentiles of this distribution.
        \end{enumerate}
        \label{enum:case-3-uncertainty}
\end{enumerate}

\begin{figure}
    \centering
    \includegraphics[width=\linewidth, keepaspectratio=false, trim=0.3cm 0.3cm 0cm 0cm,clip]{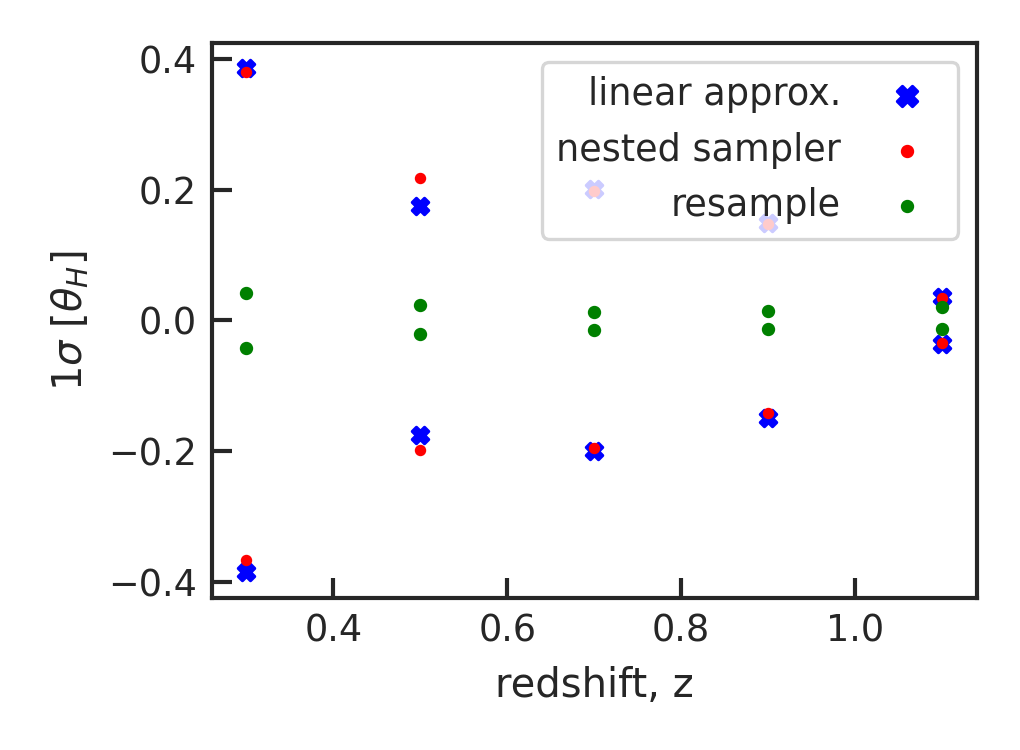}
    \caption{{Uncertainties estimated on $\theta_H$ measurements for the full area region using the different methods enumerated above: \textbf{linear approx.} - case~\ref{enum:case-1-uncertainty}, \textbf{nested sampler} - case~\ref{enum:case-2-uncertainty}, \textbf{resample} - case~\ref{enum:case-3-uncertainty}, applying the direct method for $\theta_H$ estimation. The negative and positive values correspond to the lower and upper bounds of the standard deviation.}}
    \label{fig:theta-h-uncertainties-different-methods}
\end{figure}

The resulting $1\sigma$ uncertainties for the full-area region are shown in Fig.~\ref{fig:theta-h-uncertainties-different-methods}. We find that the linear approximation (case~\ref{enum:case-1-uncertainty}) and nested sampling (case~\ref{enum:case-2-uncertainty}) approaches yield broadly consistent results, indicating that both methods capture similar features of the underlying parameter space. However, both tend to produce larger uncertainty estimates when compared to the resampling-based method (case~\ref{enum:case-3-uncertainty}).

%theta-h-uncertainties-different-methods.png

This behaviour reflects fundamental differences between the approaches. The linear approximation and nested sampling methods rely on the parametrised form of the $D_2(\theta)$ profile and propagate uncertainties from the data through the fitted model, making them biased to the assumed functional representation. In contrast, the resampling-based approach directly probes the variability of the data by accounting for spatial fluctuations within the survey footprint. 
Besides these differences, the $\theta_H$ estimated via each method are also necessarily different: in the resampling approach, it would not be computationally feasible to produce SR profiles for each $D_2$ resample, so we fit univariate splines, which make the $\theta_H$ estimates more vulnerable to small-scale oscillations from effects previously enumerated. In the nested sampling approach, the $\theta_H$ estimates are also determined by a distribution of $\theta_H$ resulting from the nested sampler samples.
Ultimately, the linear approximation method is simpler and more straightforward to implement, while not producing estimates that are too conservative, in comparison.

\subsection{Survey area dependence}

\begin{figure}
    \centering
    \includegraphics[width=\linewidth, keepaspectratio=false, trim=0.4cm 0.3cm 0.1cm 0cm,clip]{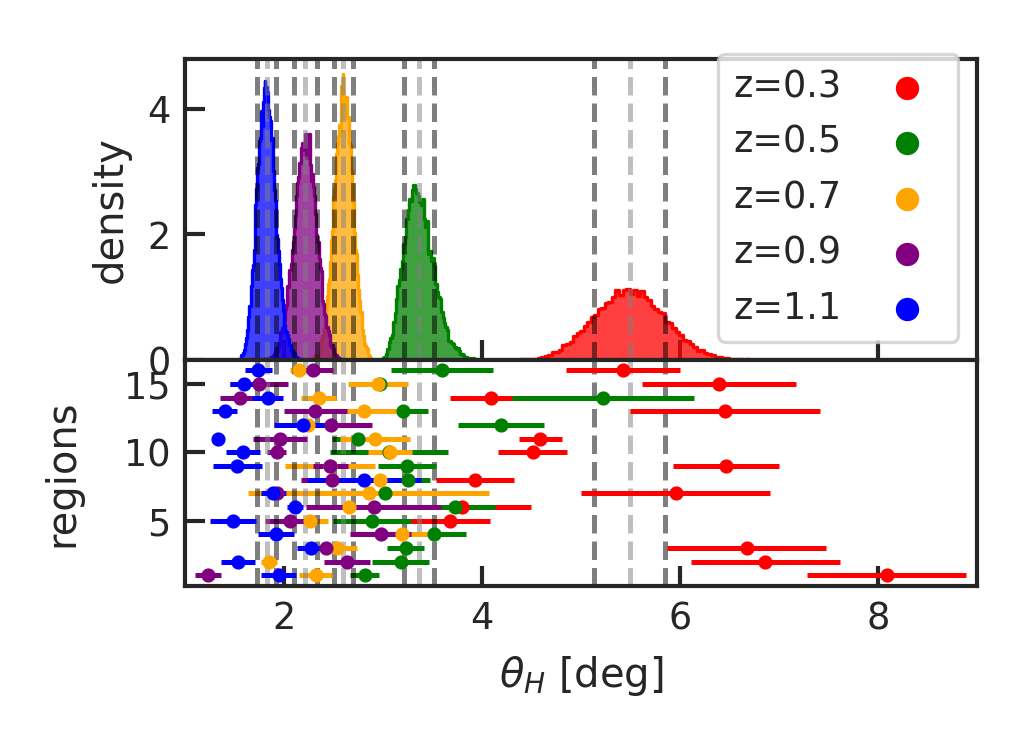} \\ 
    \vspace{-1.1cm}
    \includegraphics[width=\linewidth, keepaspectratio=false, trim=0.4cm 0.1cm 0.1cm 0.3cm,clip]{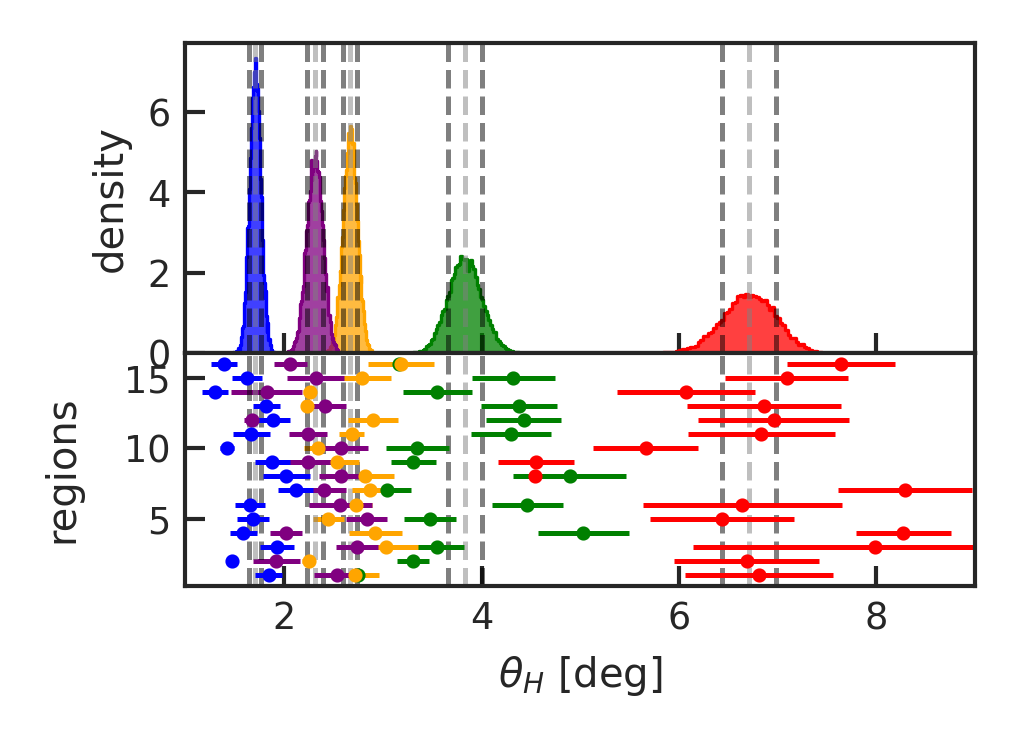}
    \caption{{\textbf{Top}: Results for regions of $600~\text{deg}^2$. \textbf{Bottom}: Results for regions of $1200~\text{deg}^2$. \\ Histograms of bootstrap distributions (density-$\theta_H$ plot) from $\theta_H$ measurements plus uncertainties (regions-$\theta_H$ plot). Results obtained using the \textbf{SR approach} with the \textbf{direct method} over many different regions of different areas. Uncertainties estimated via the \textbf{linear approximation} method.}}
    \label{fig:boot-hists-many-regions-vs-fullarea-d2-resample_non-parametric_SR}
\end{figure}

\begin{figure}
    \centering
    \includegraphics[width=\linewidth, keepaspectratio=false, trim=0.4cm 0.3cm 0.1cm 0cm,clip]{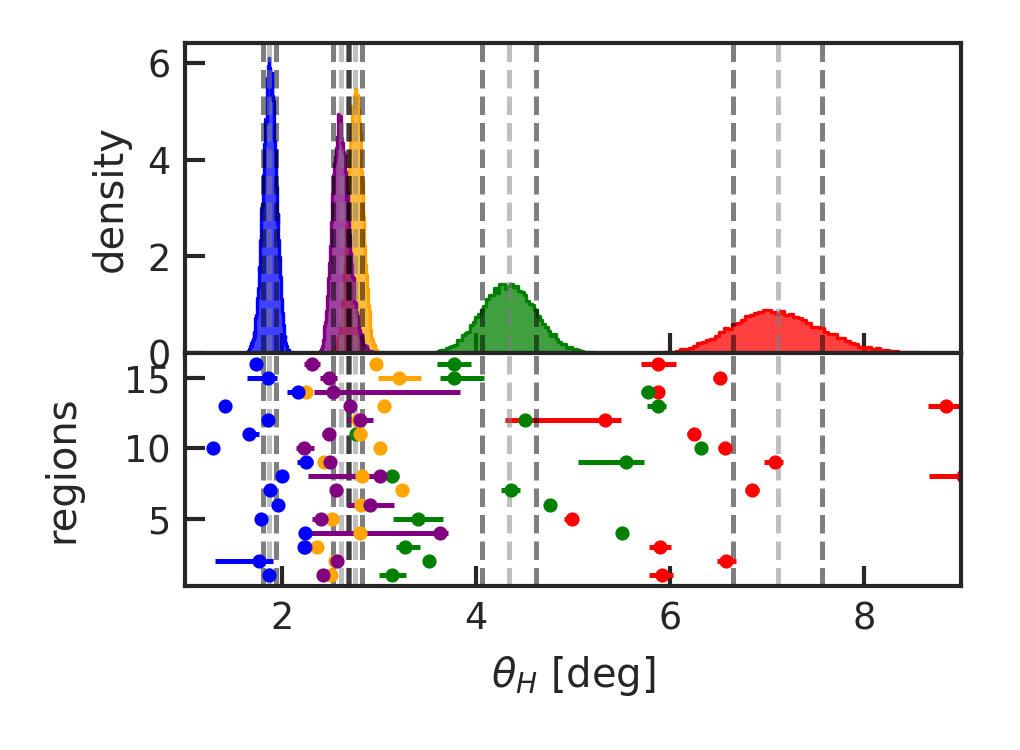} \\ 
    \vspace{-1.1cm}
    \includegraphics[width=\linewidth, keepaspectratio=false, trim=0.4cm 0.1cm 0.1cm 0.3cm,clip]{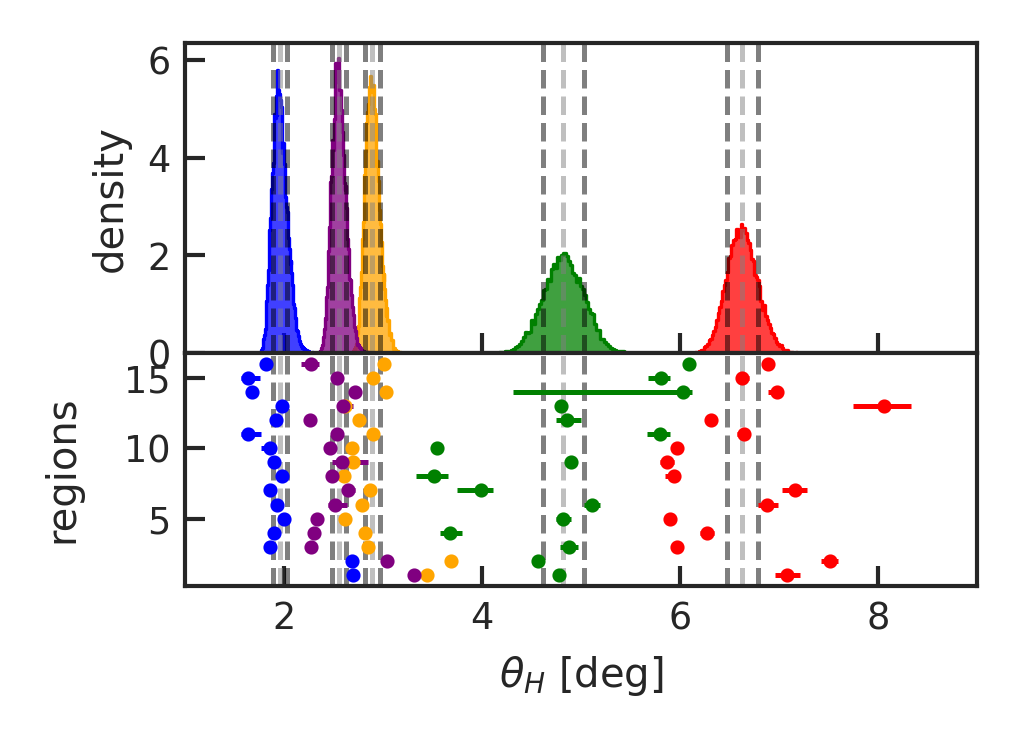} 
    \caption{{\textbf{Top}: Results for regions of $600~\text{deg}^2$. \textbf{Bottom}: Results for regions of $1200~\text{deg}^2$. \\ Histograms of bootstrap distributions (density-$\theta_H$ plot) from $\theta_H$ measurements plus uncertainties (regions-$\theta_H$ plot). Results obtained using the \textbf{resampling approach} with the \textbf{direct method} over many different regions of different areas.}}
    \label{fig:boot-hists-many-regions-vs-fullarea-d2-resample_parametric_model}
\end{figure}

In Figures~\ref{fig:boot-hists-many-regions-vs-fullarea-d2-resample_non-parametric_SR} and~\ref{fig:boot-hists-many-regions-vs-fullarea-d2-resample_parametric_model} we investigate the impact of survey area on the stability of estimation of the homogeneity scale $\theta_H$. We analyse multiple realizations of regions with areas of $600~\text{deg}^2$ (top panels) and $1200~\text{deg}^2$ (bottom panels), for both the SR approach, in Fig.~\ref{fig:boot-hists-many-regions-vs-fullarea-d2-resample_non-parametric_SR}, and for the resampling approach, in Fig.~\ref{fig:boot-hists-many-regions-vs-fullarea-d2-resample_parametric_model}. We follow the methodologies described in \ref{subsc:uncertainty-estimation} and in \citet{fanha2025angularscale}.

In both figures, in each panel, each point represents an individual measurement of $\theta_H$ obtained from a given region realization (see Fig.~\ref{fig:patches-subsampling-example}), while the associated error bars correspond to the uncertainties estimated via the linear approximation approach. The vertical dashed lines indicate the mean $\theta_H$ and the corresponding $\pm 1\sigma$ intervals inferred from bootstrapping over the sixteen region realizations within the full-area region for each redshift bin. 
The distributions shown in each panel thus provide a direct visualization of the dispersion in $\theta_H$ across different regions for a given survey area. A comparison with the corresponding full-area measurements is presented in the following section.

A clear dependence on survey area is observed in both approaches. The $600~\text{deg}^2$ regions exhibit larger scatter in $\theta_H$ values compared to the $1200~\text{deg}^2$ regions, reflecting the stronger impact of sample variance in smaller survey volumes. As the survey area increases, the distributions become more concentrated and the separation between redshift bins becomes more pronounced, indicating improved stability of the inferred homogeneity scale.

However, important differences emerge when comparing the two fitting methodologies. The SR approach shows a significantly larger dispersion in $\theta_H$ across subregions, together with broader distributions. In contrast, the resampling approach yields more compact distributions, indicating a reduced sensitivity to local fluctuations in the $D_2(\theta)$ profile. In addition, we find a bias in the $\theta_H$ estimates between the two approaches.

{We explain this as the result of the different ways in which the fitting procedures work: in the SR approach, the symbolic regression algorithm attempts to find a general expression that describes the entirety of the $D_2$ profile. Although it is given the liberty to produce an expression with large complexity, in the end we select a final expression for $D_2$ that balances complexity and quality of fit, which leads to a final fit that is less overfit than in the alternative resampling approach.}

{In this alternative approach, we fit a univariate spline that passes through every data point, such that small-scale oscillations are considered. These oscillations in the profiles can vary from region to region and have many causes, including effects from the discretization done to estimate $D_2$ as well as from baryonic acoustic oscillations (BAO).}

{In this sense, the value of $\theta_H$ determined from the SR approach is more sensitive to changes in the $D_2$ profile than when determined from the resampling approach. Nonetheless, we cannot say that the values from one approach are less valid than those from the other approach.}

It is also important to note that, due to the finite size of the simulated catalogue ($\sim 3400~\text{deg}^2$), the different realizations of subregions are not fully independent and exhibit partial overlap. This overlap is random between realizations but implies that the sampled volumes are not statistically disjoint. As a result, the observed dispersion likely underestimates the true cosmic variance expected for fully independent survey regions of the same size, representing an additional limitation associated with the finite survey area.

Overall, these results highlight the interplay between survey geometry, physical clustering scales, and methodological choices. While larger survey areas lead to more robust estimates of $\theta_H$, the comparison between fitting approaches reveals that the inferred homogeneity scale can be sensitive to the presence of BAO-scale features, particularly when analysing smaller regions. This reinforces the need to carefully assess methodological systematics when interpreting homogeneity measurements in current and future large-scale surveys.

\subsection{Redshift evolution of the homogeneity scale}

\begin{figure}
\centering
\includegraphics[width=\linewidth, keepaspectratio=false, trim=0.4cm 0.18cm 0.1cm 0cm,clip]{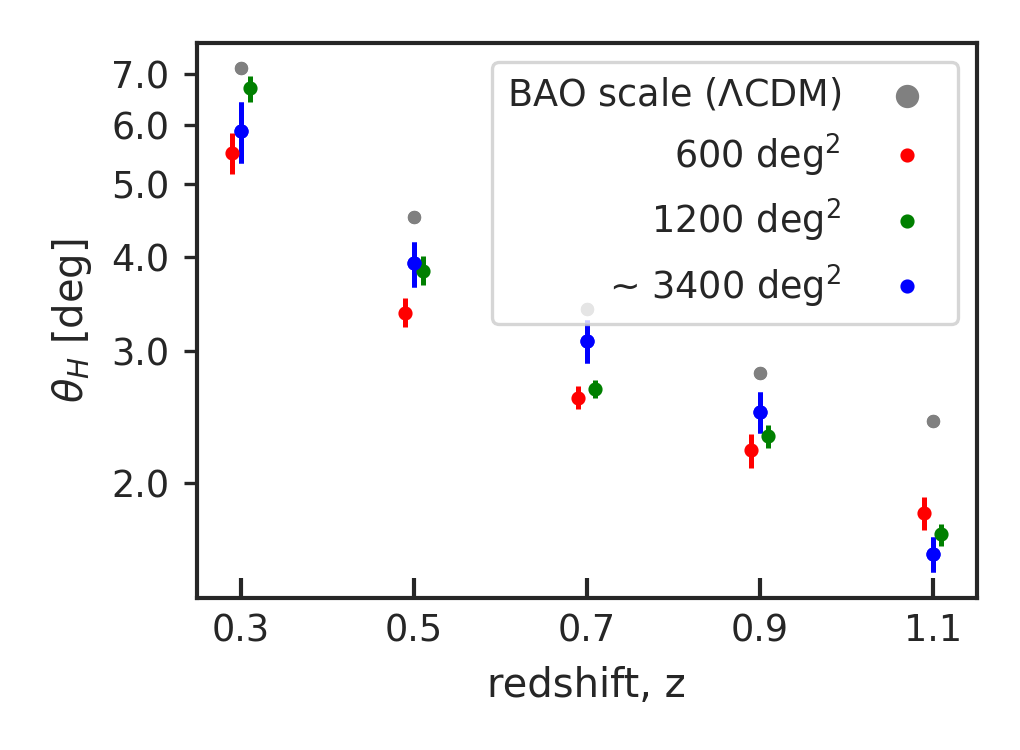} \\
\vspace{-1.1cm}
\includegraphics[width=\linewidth, keepaspectratio=false, trim=0.4cm 0.4cm 0.1cm 0.02cm]{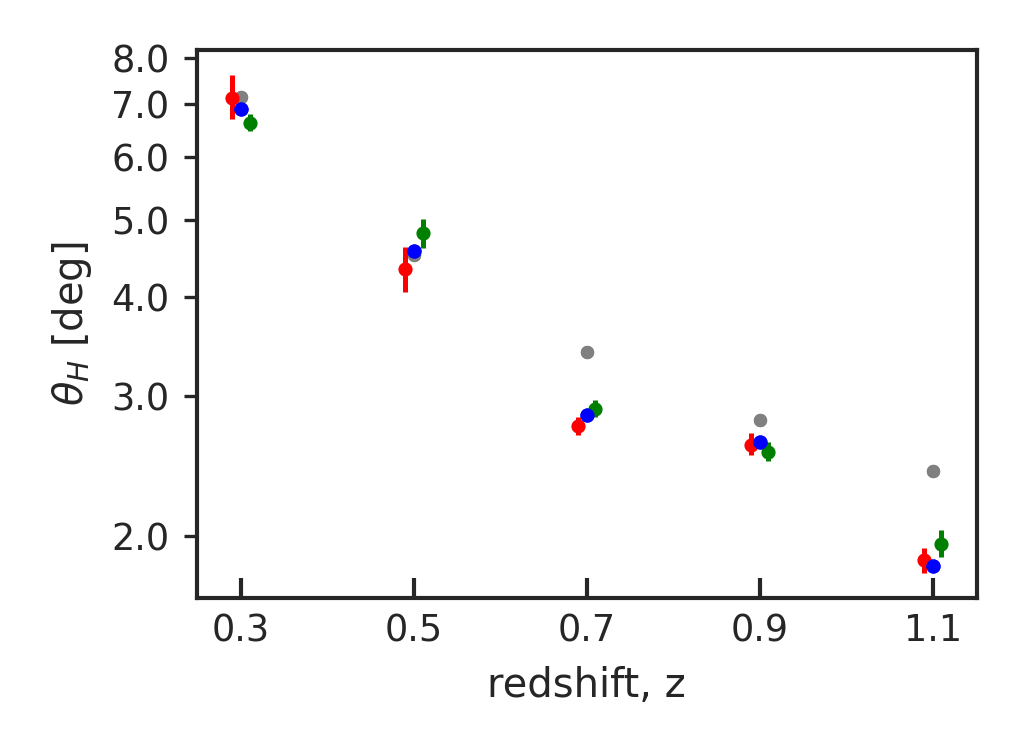}
\caption{{\textbf{Top}: Results for SR approach. \textbf{Bottom}: Results for resampling approach. \\ Evolution of the homogeneity scale $\theta_H$ as a function of redshift. Measurements from subregions of $600~\text{deg}^2$ (red) and $1200~\text{deg}^2$ (green) are compared with the estimate from the full-area region ($\sim 3400~\text{deg}^2$, blue). The grey points indicate the expected BAO angular scale computed from a fiducial $\Lambda$CDM cosmology.}}
\label{fig:theta-h-vs-z-many-regions-vs-fullarea-d2-resample-with-bao-scales}
\end{figure}

Finally, we present the measurements of the homogeneity scale, expressed as $\theta_H$, as a function of redshift for both the SR and resampling approaches, shown in Fig.~\ref{fig:theta-h-vs-z-many-regions-vs-fullarea-d2-resample-with-bao-scales}.

A clear and systematic decrease of $\theta_H$ with increasing redshift is observed in both approaches. This behaviour is expected, as a fixed physical homogeneity scale subtends a smaller angular size at higher redshift due to the evolution of the angular diameter distance. The overall trend is therefore consistent with expectations from standard cosmological models.

{The dispersion of $\theta_H$ measurements is seen to decrease towards higher redshift in both approaches, also consistent with the fact that cosmic variance decreases in that direction, thus leading to less dispersed and more consistent $D_2$ profiles.}

For both methods, the comparison between different survey areas shows that the mean values obtained from $600~\text{deg}^2$ and $1200~\text{deg}^2$ regions tend to converge towards the full-area measurement as $z$ increases. This confirms that, despite the increased dispersion observed in smaller regions, averaging over multiple subregions provides a reliable estimate of the global homogeneity scale.

{Overall, these results demonstrate that while the redshift evolution of $\theta_H$ is robust, its absolute determination can be affected by both methodological choices, redshift and the proximity to other physical scales. This highlights the importance of jointly considering modelling assumptions, survey characteristics and redshift when using $\theta_H$ as a cosmological probe.}

%%%%%%%%%%%%%%%%%%%%%%%%%%%%%%%%%%%%%%%%%%%%%%%%%%%%%%
\section{Conclusions} \label{sc:conclusions}

In this work, we have presented a systematic assessment of different computational approaches to estimate the angular homogeneity scale, $\theta_H$, using simulated galaxy catalogues from the MICE Grand Challenge. Our goal was to evaluate the robustness of $\theta_H$ measurements and to quantify the impact of methodological choices, survey geometry, and physical clustering features on its determination.

We first analysed the reconstruction of the fractal dimension profile $D_2(\theta)$ using a non-parametric symbolic regression (SR) approach applied to the direct estimator. This method provides flexible analytic representations of the data without imposing predefined functional forms, allowing the reconstructed profiles to adapt to both the strongly inhomogeneous regime and the transition to homogeneity. The resulting expressions remain compact while capturing the relevant features of the measured profiles across different redshifts.

We then investigated different strategies for estimating the uncertainty on $\theta_H$. We found that approaches based on linear error propagation and nested sampling {applied on the SR fit mathematical expressions} yield consistent results, although they tend to produce {much} more conservative uncertainties compared to a pure survey resampling-based method. The latter directly probes the variability of the data within the survey footprint and is therefore better suited to capture the impact of sample variance and survey geometry.

The dependence of $\theta_H$ on survey area was examined by analysing multiple realizations of subregions. As expected, smaller areas exhibit significantly larger dispersion due to increased sensitivity to local inhomogeneities and large-scale modes, while larger areas provide more stable and reproducible estimates. Nevertheless, averaging over multiple subregions yields mean values consistent with those obtained from the full survey area, indicating that reliable global estimates can be recovered even from smaller patches when treated statistically.

A clear and systematic decrease of $\theta_H$ with redshift was observed, consistent with the expected scaling of angular sizes with cosmic distance. This behaviour is robust across different survey areas and estimation methods, supporting the use of $\theta_H$ as a tracer of large-scale structure evolution.

{The fitting methodology selected can impact the results due to several factors in the $D_2$ profiles, including discreteness effects and the BAO. Though the two approaches did not yield results with more than $1\sigma$ deviation, one should keep this in mind when applying a methodology to future surveys with irregular footprints and masks. Discreteness effects and BAO impact may be studied in addition to minimize this difference: for instance, one should study the impact of the choice of redshift binning on the BAO clump in $D_2$.}

In general, our results show that while the redshift evolution and large-scale behavior of $\theta_H$ are robust, its absolute determination can be sensitive to both methodological choices and the presence of nearby physical scales such as the BAO feature. This highlights the importance of carefully designing analysis strategies, including redshift binning and sample selection, to minimize such effects when using $\theta_H$ as a cosmological probe.

Looking ahead, future large-scale surveys will provide significantly improved statistics and sky coverage, enabling more precise measurements of the homogeneity scale. In this context, approaches such as symbolic regression may offer a valuable tool for flexible modeling, while systematic comparisons between methods will remain essential to ensure the robustness of cosmological inferences.

\begin{acknowledgements}
 The authors are grateful to the Fundação para a Ciência e a Tecnologia (FCT) for the IA running research grant UID/04434/2025, EXPL/FIS-AST/1368/2021 (ML\_CLUSTER, DOI 10.54499/EXPL/FIS-AST/1368/2021) and PTDC/FIS-AST/0054/2021 (BEYLA, DOI 10.54499/PTDC/FIS-AST/0054/2021). JF acknowledges the support from FCT in the form of work through the Scientific Employment Incentive program (reference 2020.02633.CEECIND/CP1631/CT0002) and the FCT grant  2023.15069.PEX. ZS acknowledges support from the research projects PID2021-123012NB-C43, PID2024-159420NB-C43, the Proyecto de Investigación SAFE25003 from the Consejo Superior de Investigaciones Científicas (CSIC), and the Spanish Research Agency (Agencia Estatal de Investigaci\'on) through the Grant IFT Centro de Excelencia Severo Ochoa No CEX2020-001007-S, funded by MCIN/AEI/10.13039/501100011033.

\end{acknowledgements}

\bibliographystyle{aa}
\bibliography{references}

@ARTICLE{Carvalho_2026_closedforms,
       author = {{Carvalho}, A. and {Krone-Martins}, A. and {Da Silva}, A. and {Mimoso}, J.~P. and {B{\oe}hm}, C.},
        title = "{Closed-form approximations of fundamental quantities of Lemaitre-Tolman-Bondi cosmologies from Symbolic Regression: I. Results on the Garcia-Bellido-Haugb{\o}lle parameterization}",
      journal = {arXiv e-prints},
         year = 2026,
        month = mar,
          eid = {arXiv:2603.21277},
        pages = {arXiv:2603.21277},
          doi = {10.48550/arXiv.2603.21277},
archivePrefix = {arXiv},
       eprint = {2603.21277},
 primaryClass = {astro-ph.CO},
       adsurl = {https://ui.adsabs.harvard.edu/abs/2026arXiv260321277C}
}

@INPROCEEDINGS{Carvalho2023_sigma8,
  author={Carvalho, Ana and Oliveira, David Magalhaes and Krone-Martins, Alberto and Da Silva, Antonio},
  booktitle={2023 IEEE 19th International Conference on e-Science (e-Science)}, 
  title={Symbolic Regression Applied to Cosmology: An Approximate Expression for the Density Perturbation Variance}, 
  year={2023},
  volume={},
  number={},
  pages={1-2},
  doi={10.1109/e-Science58273.2023.10254889}
  }

@ARTICLE{Arjona2020b,
       author = {{Arjona}, Rub{\'e}n and {Nesseris}, Savvas},
        title = "{What can machine learning tell us about the background expansion of the Universe?}",
      journal = {\prd},
         year = 2020,
        month = jun,
       volume = {101},
       number = {12},
          eid = {123525},
        pages = {123525},
          doi = {10.1103/PhysRevD.101.123525},
archivePrefix = {arXiv},
       eprint = {1910.01529},
 primaryClass = {astro-ph.CO},
       adsurl = {https://ui.adsabs.harvard.edu/abs/2020PhRvD.101l3525A}
}

@ARTICLE{Nesseris2022_euclid,
       author = {{Nesseris}, S. and {Sapone}, D. and {Martinelli}, M. and {Camarena}, D. and {Marra}, V. and {Sakr}, Z. and {Garcia-Bellido}, J. and {Martins}, C.~J.~A.~P. and {Clarkson}, C. and {Da Silva}, A. and {Fleury}, P. and {Lombriser}, L. and {Mimoso}, J.~P. and {Casas}, S. and {Pettorino}, V. and {Tutusaus}, I. and {Amara}, A. and {Auricchio}, N. and {Bodendorf}, C. and {Bonino}, D. and {Branchini}, E. and {Brescia}, M. and {Capobianco}, V. and {Carbone}, C. and {Carretero}, J. and {Castellano}, M. and {Cavuoti}, S. and {Cimatti}, A. and {Cledassou}, R. and {Congedo}, G. and {Conversi}, L. and {Copin}, Y. and {Corcione}, L. and {Courbin}, F. and {Cropper}, M. and {Degaudenzi}, H. and {Douspis}, M. and {Dubath}, F. and {Duncan}, C.~A.~J. and {Dupac}, X. and {Dusini}, S. and {Ealet}, A. and {Farrens}, S. and {Fosalba}, P. and {Frailis}, M. and {Franceschi}, E. and {Fumana}, M. and {Garilli}, B. and {Gillis}, B. and {Giocoli}, C. and {Grazian}, A. and {Grupp}, F. and {Haugan}, S.~V.~H. and {Holmes}, W. and {Hormuth}, F. and {Jahnke}, K. and {Kermiche}, S. and {Kiessling}, A. and {Kitching}, T. and {K{\"u}mmel}, M. and {Kunz}, M. and {Kurki-Suonio}, H. and {Ligori}, S. and {Lilje}, P.~B. and {Lloro}, I. and {Mansutti}, O. and {Marggraf}, O. and {Markovic}, K. and {Marulli}, F. and {Massey}, R. and {Meneghetti}, M. and {Merlin}, E. and {Meylan}, G. and {Moresco}, M. and {Moscardini}, L. and {Munari}, E. and {Niemi}, S.~M. and {Padilla}, C. and {Paltani}, S. and {Pasian}, F. and {Pedersen}, K. and {Percival}, W.~J. and {Poncet}, M. and {Popa}, L. and {Racca}, G.~D. and {Raison}, F. and {Rhodes}, J. and {Roncarelli}, M. and {Saglia}, R. and {Sartoris}, B. and {Schneider}, P. and {Secroun}, A. and {Seidel}, G. and {Serrano}, S. and {Sirignano}, C. and {Sirri}, G. and {Stanco}, L. and {Starck}, J.-L. and {Tallada-Cresp{\'\i}}, P. and {Taylor}, A.~N. and {Tereno}, I. and {Toledo-Moreo}, R. and {Torradeflot}, F. and {Valentijn}, E.~A. and {Valenziano}, L. and {Wang}, Y. and {Welikala}, N. and {Zamorani}, G. and {Zoubian}, J. and {Andreon}, S. and {Baldi}, M. and {Camera}, S. and {Medinaceli}, E. and {Mei}, S. and {Renzi}, A.},
        title = "{Euclid: Forecast constraints on consistency tests of the {\ensuremath{\Lambda}}CDM model}",
      journal = {\aap},
         year = 2022,
        month = apr,
       volume = {660},
          eid = {A67},
        pages = {A67},
          doi = {10.1051/0004-6361/202142503},
archivePrefix = {arXiv},
       eprint = {2110.11421},
 primaryClass = {astro-ph.CO},
       adsurl = {https://ui.adsabs.harvard.edu/abs/2022A&A...660A..67N}
}

@misc{shao2025cosmologicalconstraintsangularhomogeneity,
      title={Cosmological constraints from angular homogeneity scale measurements}, 
      author={Xiaoyun Shao and Carlos A. P. Bengaly and Rodrigo S. Gonçalves and Gabriela C. Carvalho and Jailson Alcaniz},
      year={2025},
      eprint={2409.06009},
      archivePrefix={arXiv},
      primaryClass={astro-ph.CO},
      url={https://arxiv.org/abs/2409.06009}, 
}

@article{Avila_2019,
   title={The angular scale of homogeneity in the local Universe with the SDSS blue galaxies},
   volume={488},
   ISSN={1365-2966},
   url={http://dx.doi.org/10.1093/mnras/stz1765},
   DOI={10.1093/mnras/stz1765},
   number={1},
   journal={Monthly Notices of the Royal Astronomical Society},
   publisher={Oxford University Press (OUP)},
   author={Avila, F and Novaes, C P and Bernui, A and de Carvalho, E and Nogueira-Cavalcante, J P},
   year={2019},
   month=jul, pages={1481–1487} }

@ARTICLE{Aluri_2023,
       author = {{Aluri}, Pavan kumar and {Cea}, Paolo and {Chingangbam}, Pravabati and {Chu}, Ming-Chung and {Clowes}, Roger G. and {Hutsem{\'e}kers}, Damien and {Kochappan}, Joby P. and {Lopez}, Alexia M. and {Liu}, Lang and {Martens}, Niels C.~M. and {Martins}, C.~J.~A.~P. and {Migkas}, Konstantinos and {{\'O} Colg{\'a}in}, Eoin and {Pranav}, Pratyush and {Shamir}, Lior and {Singal}, Ashok K. and {Sheikh-Jabbari}, M.~M. and {Wagner}, Jenny and {Wang}, Shao-Jiang and {Wiltshire}, David L. and {Yeung}, Shek and {Yin}, Lu and {Zhao}, Wen},
        title = "{Is the observable Universe consistent with the cosmological principle?}",
      journal = {Classical and Quantum Gravity},
         year = 2023,
        month = may,
       volume = {40},
       number = {9},
          eid = {094001},
        pages = {094001},
          doi = {10.1088/1361-6382/acbefc},
archivePrefix = {arXiv},
       eprint = {2207.05765},
 primaryClass = {astro-ph.CO},
       adsurl = {https://ui.adsabs.harvard.edu/abs/2023CQGra..40i4001A}
}

@ARTICLE{Goncalves_2018,
       author = {{Gon{\c{c}}alves}, R.~S. and {Carvalho}, G.~C. and {Bengaly}, Jr., C.~A.~P. and {Carvalho}, J.~C. and {Bernui}, A. and {Alcaniz}, J.~S. and {Maartens}, R.},
        title = "{Cosmic homogeneity: a spectroscopic and model-independent measurement}",
      journal = {MNRAS},
         year = 2018,
        month = mar,
       volume = {475},
       number = {1},
        pages = {L20-L24},
          doi = {10.1093/mnrasl/slx202},
archivePrefix = {arXiv},
       eprint = {1710.02496},
 primaryClass = {astro-ph.CO},
       adsurl = {https://ui.adsabs.harvard.edu/abs/2018MNRAS.475L..20G}
}

@ARTICLE{Shao_2025,
       author = {{Shao}, Xiaoyun and {Bengaly}, Carlos A.~P. and {Gon{\c{c}}alves}, Rodrigo S. and {Carvalho}, Gabriela C. and {Alcaniz}, Jailson},
        title = "{Cosmological constraints from angular homogeneity scale measurements}",
      journal = {European Physical Journal C},
         year = 2025,
        month = mar,
       volume = {85},
       number = {3},
          eid = {225},
        pages = {225},
          doi = {10.1140/epjc/s10052-025-13987-4},
archivePrefix = {arXiv},
       eprint = {2409.06009},
 primaryClass = {astro-ph.CO},
       adsurl = {https://ui.adsabs.harvard.edu/abs/2025EPJC...85..225S}
}

@article{Gon_alves_2017,
   title={Cosmic homogeneity: a spectroscopic and model-independent measurement},
   volume={475},
   ISSN={1745-3933},
   url={http://dx.doi.org/10.1093/mnrasl/slx202},
   DOI={10.1093/mnrasl/slx202},
   number={1},
   journal={Monthly Notices of the Royal Astronomical Society: Letters},
   publisher={Oxford University Press (OUP)},
   author={Gonçalves, R S and Carvalho, G C and Bengaly Jr, C A P and Carvalho, J C and Bernui, A and Alcaniz, J S and Maartens, R},
   year={2017},
   month=dec, pages={L20–L24} }

@article{Avila_2018,
   title={The scale of homogeneity in the local Universe with the ALFALFA catalogue},
   volume={2018},
   ISSN={1475-7516},
   url={http://dx.doi.org/10.1088/1475-7516/2018/12/041},
   DOI={10.1088/1475-7516/2018/12/041},
   number={12},
   journal={Journal of Cosmology and Astroparticle Physics},
   publisher={IOP Publishing},
   author={Avila, Felipe and Novaes, Camila P. and Bernui, Armando and de Carvalho, Edilson},
   year={2018},
   month=dec, pages={041–041} }

@misc{andrade2022angularscalehomogeneitysdssiv,
      title={The angular scale of homogeneity with SDSS-IV DR16 Luminous Red Galaxies}, 
      author={Uendert Andrade and Rodrigo S. Gonçalves and Gabriela C. Carvalho and Carlos A. P. Bengaly and Joel C. Carvalho and Jailson Alcaniz},
      year={2022},
      eprint={2205.07819},
      archivePrefix={arXiv},
      primaryClass={astro-ph.CO},
      url={https://arxiv.org/abs/2205.07819}, 
}

@article{Mohammad_2022,
   title={Creating jackknife and bootstrap estimates of the covariance matrix for the two-point correlation function},
   volume={514},
   ISSN={1365-2966},
   url={http://dx.doi.org/10.1093/mnras/stac1458},
   DOI={10.1093/mnras/stac1458},
   number={1},
   journal={Monthly Notices of the Royal Astronomical Society},
   publisher={Oxford University Press (OUP)},
   author={Mohammad, Faizan G and Percival, Will J},
   year={2022},
   month=may, pages={1289–1301} }

@article{Sanchez_2011,
    author = {Sánchez, E. and Carnero, A. and García-Bellido, J. and Gaztañaga, E. and de Simoni, F. and Crocce, M. and Cabré, A. and Fosalba, P. and Alonso, D.},
    title = {Tracing the sound horizon scale with photometric redshift surveys},
    journal = {Monthly Notices of the Royal Astronomical Society},
    volume = {411},
    number = {1},
    pages = {277-288},
    year = {2011},
    month = {01},
    issn = {0035-8711},
    doi = {10.1111/j.1365-2966.2010.17679.x},
    url = {https://doi.org/10.1111/j.1365-2966.2010.17679.x},
    eprint = {https://academic.oup.com/mnras/article-pdf/411/1/277/3509047/mnras0411-0277.pdf},
}

@article{Ntelis_2017,
   title={Exploring cosmic homogeneity with the BOSS DR12 galaxy sample},
   volume={2017},
   ISSN={1475-7516},
   url={http://dx.doi.org/10.1088/1475-7516/2017/06/019},
   DOI={10.1088/1475-7516/2017/06/019},
   number={06},
   journal={Journal of Cosmology and Astroparticle Physics},
   publisher={IOP Publishing},
   author={Ntelis, Pierros and Hamilton, Jean-Christophe and Goff, Jean-Marc Le and Burtin, Etienne and Laurent, Pierre and Rich, James and Busca, Nicolas Guillermo and Tinker, Jeremy and Aubourg, Eric and Bourboux, Hélion du Mas des and Bautista, Julian and Delabrouille, Nathalie Palanque and Delubac, Timothée and Eftekharzadeh, Sarah and Hogg, David W. and Myers, Adam and Vargas-Magaña, Mariana and Pâris, Isabelle and Petitjean, Partick and Rossi, Graziano and Schneider, Donald P. and Tojeiro, Rita and Yeche, Christophe},
   year={2017},
   month=jun, pages={019–019} }

@article{Scrimgeour_2012,
   title={The WiggleZ Dark Energy Survey: the transition to large-scale cosmic homogeneity: Cosmic homogeneity in the WiggleZ survey},
   volume={425},
   ISSN={0035-8711},
   url={http://dx.doi.org/10.1111/j.1365-2966.2012.21402.x},
   DOI={10.1111/j.1365-2966.2012.21402.x},
   number={1},
   journal={Monthly Notices of the Royal Astronomical Society},
   publisher={Oxford University Press (OUP)},
   author={Scrimgeour, Morag I. and Davis, Tamara and Blake, Chris and James, J. Berian and Poole, Gregory B. and Staveley-Smith, Lister and Brough, Sarah and Colless, Matthew and Contreras, Carlos and Couch, Warrick and Croom, Scott and Croton, Darren and Drinkwater, Michael J. and Forster, Karl and Gilbank, David and Gladders, Mike and Glazebrook, Karl and Jelliffe, Ben and Jurek, Russell J. and Li, I-hui and Madore, Barry and Martin, D. Christopher and Pimbblet, Kevin and Pracy, Michael and Sharp, Rob and Wisnioski, Emily and Woods, David and Wyder, Ted K. and Yee, H. K. C.},
   year={2012},
   month=jul, pages={116–134} }

@misc{dias2023probingcosmichomogeneitylocal,
      title={Probing cosmic homogeneity in the Local Universe}, 
      author={Bruno L. Dias and Felipe Avila and Armando Bernui},
      year={2023},
      eprint={2310.04594},
      archivePrefix={arXiv},
      primaryClass={astro-ph.CO},
      url={https://arxiv.org/abs/2310.04594}, 
}

@article{Hogg_2005,
   title={Cosmic Homogeneity Demonstrated with Luminous Red Galaxies},
   volume={624},
   ISSN={1538-4357},
   url={http://dx.doi.org/10.1086/429084},
   DOI={10.1086/429084},
   number={1},
   journal={The Astrophysical Journal},
   publisher={American Astronomical Society},
   author={Hogg, David W. and Eisenstein, Daniel J. and Blanton, Michael R. and Bahcall, Neta A. and Brinkmann, J. and Gunn, James E. and Schneider, Donald P.},
   year={2005},
   month=may, pages={54–58} }

@ARTICLE{Abbot_2021,
       author = {{Abbott}, T.~M.~C. and {Adam{\'o}w}, M. and {Aguena}, M. and {Allam}, S. and {Amon}, A. and {Annis}, J. and {Avila}, S. and {Bacon}, D. and {Banerji}, M. and {Bechtol}, K. and {Becker}, M.~R. and {Bernstein}, G.~M. and {Bertin}, E. and {Bhargava}, S. and {Bridle}, S.~L. and {Brooks}, D. and {Burke}, D.~L. and {Carnero Rosell}, A. and {Carrasco Kind}, M. and {Carretero}, J. and {Castander}, F.~J. and {Cawthon}, R. and {Chang}, C. and {Choi}, A. and {Conselice}, C. and {Costanzi}, M. and {Crocce}, M. and {da Costa}, L.~N. and {Davis}, T.~M. and {De Vicente}, J. and {DeRose}, J. and {Desai}, S. and {Diehl}, H.~T. and {Dietrich}, J.~P. and {Drlica-Wagner}, A. and {Eckert}, K. and {Elvin-Poole}, J. and {Everett}, S. and {Evrard}, A.~E. and {Ferrero}, I. and {Fert{\'e}}, A. and {Flaugher}, B. and {Fosalba}, P. and {Friedel}, D. and {Frieman}, J. and {Garc{\'\i}a-Bellido}, J. and {Gaztanaga}, E. and {Gelman}, L. and {Gerdes}, D.~W. and {Giannantonio}, T. and {Gill}, M.~S.~S. and {Gruen}, D. and {Gruendl}, R.~A. and {Gschwend}, J. and {Gutierrez}, G. and {Hartley}, W.~G. and {Hinton}, S.~R. and {Hollowood}, D.~L. and {Honscheid}, K. and {Huterer}, D. and {James}, D.~J. and {Jeltema}, T. and {Johnson}, M.~D. and {Kent}, S. and {Kron}, R. and {Kuehn}, K. and {Kuropatkin}, N. and {Lahav}, O. and {Li}, T.~S. and {Lidman}, C. and {Lin}, H. and {MacCrann}, N. and {Maia}, M.~A.~G. and {Manning}, T.~A. and {Maloney}, J.~D. and {March}, M. and {Marshall}, J.~L. and {Martini}, P. and {Melchior}, P. and {Menanteau}, F. and {Miquel}, R. and {Morgan}, R. and {Myles}, J. and {Neilsen}, E. and {Ogando}, R.~L.~C. and {Palmese}, A. and {Paz-Chinch{\'o}n}, F. and {Petravick}, D. and {Pieres}, A. and {Plazas}, A.~A. and {Pond}, C. and {Rodriguez-Monroy}, M. and {Romer}, A.~K. and {Roodman}, A. and {Rykoff}, E.~S. and {Sako}, M. and {Sanchez}, E. and {Santiago}, B. and {Scarpine}, V. and {Serrano}, S. and {Sevilla-Noarbe}, I. and {Smith}, J. Allyn and {Smith}, M. and {Soares-Santos}, M. and {Suchyta}, E. and {Swanson}, M.~E.~C. and {Tarle}, G. and {Thomas}, D. and {To}, C. and {Tremblay}, P.~E. and {Troxel}, M.~A. and {Tucker}, D.~L. and {Turner}, D.~J. and {Varga}, T.~N. and {Walker}, A.~R. and {Wechsler}, R.~H. and {Weller}, J. and {Wester}, W. and {Wilkinson}, R.~D. and {Yanny}, B. and {Zhang}, Y. and {Nikutta}, R. and {Fitzpatrick}, M. and {Jacques}, A. and {Scott}, A. and {Olsen}, K. and {Huang}, L. and {Herrera}, D. and {Juneau}, S. and {Nidever}, D. and {Weaver}, B.~A. and {Adean}, C. and {Correia}, V. and {de Freitas}, M. and {Freitas}, F.~N. and {Singulani}, C. and {Vila-Verde}, G. and {Linea Science Server}},
        title = "{The Dark Energy Survey Data Release 2}",
      journal = {Astrophysical Journal, Supplement Series},
         year = 2021,
        month = aug,
       volume = {255},
       number = {2},
          eid = {20},
        pages = {20},
          doi = {10.3847/1538-4365/ac00b3},
archivePrefix = {arXiv},
       eprint = {2101.05765},
 primaryClass = {astro-ph.IM},
       adsurl = {https://ui.adsabs.harvard.edu/abs/2021ApJS..255...20A}
}

@ARTICLE{Scaramella-EP1,
       author = {{Euclid Collaboration: Scaramella}, R. and {Amiaux}, J. and {Mellier}, Y. and others},
        title = "{Euclid preparation. I. The Euclid Wide Survey}",
      journal = {Astronomy and Astrophysics},
         year = 2022,
        month = jun,
       volume = {662},
          eid = {A112},
        pages = {A112},
          doi = {10.1051/0004-6361/202141938},
archivePrefix = {arXiv},
       eprint = {2108.01201},
 primaryClass = {astro-ph.CO},
       adsurl = {https://ui.adsabs.harvard.edu/abs/2022A&A...662A.112E}
}

@ARTICLE{Fosalba_2015a,
       author = {{Fosalba}, P. and {Crocce}, M. and {Gazta{\~n}aga}, E. and {Castander},
        F.~J.},
        title = "{The MICE grand challenge lightcone simulation - I. Dark matter
        clustering}",
      journal = {Monthly Notices of the Royal Astronomical Society},
         year = 2015,
        month = Apr,
       volume = {448},
        pages = {2987-3000},
          doi = {10.1093/mnras/stv138},
archivePrefix = {arXiv},
       eprint = {1312.1707},
 primaryClass = {astro-ph.CO},
       adsurl = {https://ui.adsabs.harvard.edu/\#abs/2015MNRAS.448.2987F}
}

@ARTICLE{Crocce_2015,
       author = {{Crocce}, M. and {Castander}, F.~J. and {Gazta{\~n}aga}, E. and
        {Fosalba}, P. and {Carretero}, J.},
        title = "{The MICE Grand Challenge lightcone simulation - II. Halo and galaxy
        catalogues}",
      journal = {Monthly Notices of the Royal Astronomical Society},
         year = 2015,
        month = Oct,
       volume = {453},
        pages = {1513-1530},
          doi = {10.1093/mnras/stv1708},
archivePrefix = {arXiv},
       eprint = {1312.2013},
 primaryClass = {astro-ph.CO},
       adsurl = {https://ui.adsabs.harvard.edu/\#abs/2015MNRAS.453.1513C}
}

@ARTICLE{Fosalba_2015b,
       author = {{Fosalba}, P. and {Gazta{\~n}aga}, E. and {Castander}, F.~J. and
        {Crocce}, M.},
        title = "{The MICE Grand Challenge light-cone simulation - III. Galaxy lensing
        mocks from all-sky lensing maps}",
      journal = {Monthly Notices of the Royal Astronomical Society},
         year = 2015,
        month = Feb,
       volume = {447},
        pages = {1319-1332},
          doi = {10.1093/mnras/stu2464},
archivePrefix = {arXiv},
       eprint = {1312.2947},
 primaryClass = {astro-ph.CO},
       adsurl = {https://ui.adsabs.harvard.edu/\#abs/2015MNRAS.447.1319F}
}

@INPROCEEDINGS{Carretero_2017,
       author = {{Carretero}, J. and {Tallada}, P. and {Casals}, J. and {Caubet}, M. and
        {Castander}, F. and {Blot}, L. and {Alarc{\'o}n}, A. and
        {Serrano}, S. and {Fosalba}, P. and {Acosta-Silva}, C. and
        {Tonello}, N. and {Torradeflot}, F. n. and {Eriksen}, M. and
        {Neissner}, C. and {Delfino}, M.},
        title = "{CosmoHub and SciPIC: Massive cosmological data analysis, distribution and generation using a Big Data platform}",
    booktitle = {Proceedings of the European Physical Society Conference on High Energy Physics. 5-12 July},
         year = 2017,
        month = Jul,
          eid = {488},
        pages = {488},
       adsurl = {https://ui.adsabs.harvard.edu/\#abs/2017ehep.confE.488C}
}

@ARTICLE{Aizpuru_2021,
       author = {{Aizpuru}, Andoni and {Arjona}, Rub{\'e}n and {Nesseris}, Savvas},
        title = "{Machine learning improved fits of the sound horizon at the baryon drag epoch}",
      journal = {\prd},
         year = 2021,
        month = aug,
       volume = {104},
       number = {4},
          eid = {043521},
        pages = {043521},
          doi = {10.1103/PhysRevD.104.043521},
archivePrefix = {arXiv},
       eprint = {2106.00428},
 primaryClass = {astro-ph.CO},
       adsurl = {https://ui.adsabs.harvard.edu/abs/2021PhRvD.104d3521A}
}

@book{10.23943/princeton/9780691209814.001.0001,
    author = {Peebles, P. J. E.},
    title = {Principles of Physical Cosmology},
    publisher = {Princeton University Press},
    year = {2020},
    month = {09},
    isbn = {9780691209814},
    doi = {10.23943/princeton/9780691209814.001.0001},
    url = {https://doi.org/10.23943/princeton/9780691209814.001.0001},
}

@article{Heinesen_2020,
   title={Cosmological homogeneity scale estimates are dressed},
   volume={2020},
   ISSN={1475-7516},
   url={http://dx.doi.org/10.1088/1475-7516/2020/10/052},
   DOI={10.1088/1475-7516/2020/10/052},
   number={10},
   journal={Journal of Cosmology and Astroparticle Physics},
   publisher={IOP Publishing},
   author={Heinesen, Asta},
   year={2020},
   month=oct, pages={052–052} }

@misc{Hirata2009_CF,
    author    = {Christopher M. Hirata},
    title     = {Correlation functions: Generalities},
    year      = {2009},
    note      = {Draft: November 19, 2009, Caltech cosmology notes.},
    url       = {http://www.tapir.caltech.edu/~chirata/ay211/Hirata_CF.pdf}
}

@article{Jing_2002,
   title={Triaxial Modeling of Halo Density Profiles with High‐ResolutionN‐Body Simulations},
   volume={574},
   ISSN={1538-4357},
   url={http://dx.doi.org/10.1086/341065},
   DOI={10.1086/341065},
   number={2},
   journal={The Astrophysical Journal},
   publisher={American Astronomical Society},
   author={Jing, Y. P. and Suto, Yasushi},
   year={2002},
   month=aug, pages={538–553} }

@ARTICLE{2001ApJ...546...20S,
       author = {{Scoccimarro}, Rom{\'a}n and {Sheth}, Ravi K. and {Hui}, Lam and {Jain}, Bhuvnesh},
        title = "{How Many Galaxies Fit in a Halo? Constraints on Galaxy Formation Efficiency from Spatial Clustering}",
      journal = {The Astrophysical Journal},
         year = 2001,
        month = jan,
       volume = {546},
       number = {1},
        pages = {20-34},
          doi = {10.1086/318261},
archivePrefix = {arXiv},
       eprint = {astro-ph/0006319},
 primaryClass = {astro-ph},
       adsurl = {https://ui.adsabs.harvard.edu/abs/2001ApJ...546...20S}
}

@article{Berlind_2002,
doi = {10.1086/341469},
url = {https://doi.org/10.1086/341469},
year = {2002},
month = {aug},
publisher = {},
volume = {575},
number = {2},
pages = {587},
author = {Berlind, Andreas A. and Weinberg, David H.},
title = {The Halo Occupation Distribution: Toward an Empirical Determination of the Relation between Galaxies and Mass},
journal = {The Astrophysical Journal}
}

@ARTICLE{2004MNRAS.353..189V,
       author = {{Vale}, A. and {Ostriker}, J.~P.},
        title = "{Linking halo mass to galaxy luminosity}",
      journal = {Monthly Notices of the Royal Astronomical Society},
         year = 2004,
        month = sep,
       volume = {353},
       number = {1},
        pages = {189-200},
          doi = {10.1111/j.1365-2966.2004.08059.x},
archivePrefix = {arXiv},
       eprint = {astro-ph/0402500},
 primaryClass = {astro-ph},
       adsurl = {https://ui.adsabs.harvard.edu/abs/2004MNRAS.353..189V}
}

@article{Tasitsiomi_2004,
doi = {10.1086/423784},
url = {https://doi.org/10.1086/423784},
year = {2004},
month = {oct},
publisher = {},
volume = {614},
number = {2},
pages = {533},
author = {Tasitsiomi, Argyro and Kravtsov, Andrey V. and Wechsler, Risa H. and Primack, Joel R.},
title = {Modeling Galaxy-Mass Correlations in Dissipationless Simulations},
journal = {The Astrophysical Journal}
}

@ARTICLE{2006ApJ...647..201C,
       author = {{Conroy}, Charlie and {Wechsler}, Risa H. and {Kravtsov}, Andrey V.},
        title = "{Modeling Luminosity-dependent Galaxy Clustering through Cosmic Time}",
      journal = {The Astrophysical Journal},
         year = 2006,
        month = aug,
       volume = {647},
       number = {1},
        pages = {201-214},
          doi = {10.1086/503602},
archivePrefix = {arXiv},
       eprint = {astro-ph/0512234},
 primaryClass = {astro-ph},
       adsurl = {https://ui.adsabs.harvard.edu/abs/2006ApJ...647..201C}
}

@ARTICLE{1982ApJ...257..423H,
       author = {{Huchra}, J.~P. and {Geller}, M.~J.},
        title = "{Groups of Galaxies. I. Nearby groups}",
      journal = {The Astrophysical Journal},
         year = 1982,
        month = jun,
       volume = {257},
        pages = {423-437},
          doi = {10.1086/160000},
       adsurl = {https://ui.adsabs.harvard.edu/abs/1982ApJ...257..423H}
}

@ARTICLE{1984MNRAS.206..559T,
       author = {{Tago}, E. and {Einasto}, J. and {Saar}, E.},
        title = "{Structure of superclusters and supercluster formation - IV. Spatial distribution of clusters of galaxies in the Coma Supercluster and its large-scale environment.}",
      journal = {Monthly Notices of the Royal Astronomical Society},
         year = 1984,
        month = feb,
       volume = {206},
        pages = {559-587},
          doi = {10.1093/mnras/206.3.559},
       adsurl = {https://ui.adsabs.harvard.edu/abs/1984MNRAS.206..559T}
}

@ARTICLE{1982ApJ...259..449P,
       author = {{Press}, W.~H. and {Davis}, M.},
        title = "{How to identify and weigh virialized clusters of galaxies in a complete redshift catalog}",
      journal = {The Astrophysical Journal},
         year = 1982,
        month = aug,
       volume = {259},
        pages = {449-473},
          doi = {10.1086/160183},
       adsurl = {https://ui.adsabs.harvard.edu/abs/1982ApJ...259..449P}
}

@mastersthesis{fanha2025angularscale,
  author  = {Pedro Miguel Silva Fanha},
  title   = {Testing the FLRW model with large-scale structure estimators},
  school  = {University of Porto},
  year    = {2025},
  address = {Porto, Portugal},
  type    = {Master's Dissertation},
  url     = {https://hdl.handle.net/10216/173424},
  note    = {Accessed: 2026-03-28}
}

\end{document}